\documentclass[reprint,floatfix,amsmath,amssymb,aps,superscriptaddress]{revtex4-2}

\usepackage{graphicx}
\usepackage{dcolumn}
\usepackage{bm}

\usepackage{color}

\usepackage{array}
\newcolumntype{C}[1]{>{\centering\arraybackslash}p{#1}}
\usepackage[colorlinks=true,linkcolor=blue,citecolor=cyan,urlcolor=blue]{hyperref}%

\begin{document}
\preprint{APS/123-QED}


\title{Ternary Wide Band Gap Oxides for High-Power Electronics\\Identified Computationally}

\author{Emily Garrity}
\affiliation{Colorado School of Mines, Golden, Colorado 80401, USA}
\affiliation{National Renewable Energy Laboratory (NREL), Golden, Colorado 80401, USA}

\author{Cheng-Wei Lee}
\affiliation{Colorado School of Mines, Golden, Colorado 80401, USA}
\affiliation{National Renewable Energy Laboratory (NREL), Golden, Colorado 80401, USA}

\author{Prashun Gorai}
\affiliation{Colorado School of Mines, Golden, Colorado 80401, USA}
\affiliation{National Renewable Energy Laboratory (NREL), Golden, Colorado 80401, USA}

\author{Brooks Tellekamp}
\affiliation{National Renewable Energy Laboratory (NREL), Golden, Colorado 80401, USA}

\author{Andriy Zakutayev}
\affiliation{National Renewable Energy Laboratory (NREL), Golden, Colorado 80401, USA}

\author{Vladan Stevanovi\'{c}}
\email{vstevano@mines.edu}
\affiliation{Colorado School of Mines, Golden, Colorado 80401, USA}
\affiliation{National Renewable Energy Laboratory (NREL), Golden, Colorado 80401, USA}


\begin{abstract}

As electricity grids become more renewable energy-compliant, there will be a need for novel semiconductors that can withstand high power, high voltage, and high temperatures. Wide band gap (WBG) semiconductors tend to exhibit large breakdown field, allowing high operating voltages. Currently explored WBG materials for power electronics are costly (GaN), difficult to synthesize as high-quality single crystals (SiC) and at scale (diamond, BN), have low thermal conductivity ($\beta$-Ga$_2$O$_3$), or cannot be suitably doped (AlN). We conduct a computational search for novel semiconductors across 1,340 known metal-oxides using first-principles calculations and existing transport models. We calculate the Baliga figure of merit (BFOM) and lattice thermal conductivity ($\kappa_L$) to identify top candidates for \textit{n}-type power electronics. We find 40 mostly ternary oxides that have higher $\kappa_L$ than $\beta$-Ga$_2$O$_3$ and higher \textit{n}-type BFOM than SiC and GaN. Among these, several material classes emerge, including 2-2-7 stoichiometry thortveitites and pyrochlores, II-IV spinels, and calcite-type borates. Within these classes, we propose In$_2$Ge$_2$O$_7$, Mg$_2$GeO$_4$, and InBO$_3$ as they are the most favorable for \textit{n}-type doping based on our preliminary evaluation and could be grown as single crystals or thin film heterostructures. These materials could help advance power electronic devices for the future grid.



\end{abstract}


\maketitle
%
%
%

\section{Introduction}
%
Increased electrification, smart grid technology, and renewable power generation has brought to light the need for improved power electronics. These are required for an electrical grid which is more sustainable, flexible, reliable, and distributed  \cite{kaplar_generation-after-next_2017, shinde_role_2009}. As technologies like photovoltaics and electric vehicles become more common, the electronics for power conversion (DC-AC, AC-DC) must be able to handle larger power in order to control utility-scale energy \cite{huang_power_2017}. In addition, traditionally analog technologies like transformers used in power transmission, can be replaced by solid state inverters which allow the quick transfer of distributed, renewable energy across long geographical distances \cite{huang_power_2017}. These applications require power electronics that can handle high voltages with great efficiency while keeping device size relatively small \cite{tolbert_power_2005}.

As a consequence of high power (kW to GW and more) and high voltage ($>$1200 V per component) requirements, any new power electronic device must also need to withstand elevated temperatures 
\cite{pearton_perspective_2018, huang_power_2017}. One can address this need for better devices by either engineering new architectures around existing 
semiconductors or replacing the semiconductor with an inherently higher-performing material. In this paper we focus on the latter: the identification of previously unexplored semiconductors for future power electronics.

\begin{figure*}[!t]
    \includegraphics[width=0.65\textwidth]{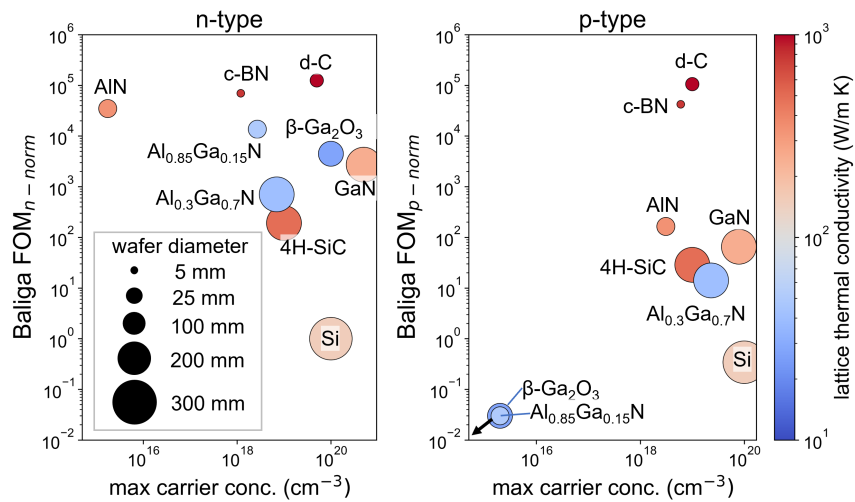}
    \caption{ \label{fig:motivation} \textit{N}-type (left) and \textit{p}-type (right) performance of commercialized and exploratory power electronic semiconductors in terms of the Baliga figure of merit (relative to \textit{n}-type Si), thermal conductivity, and 
    maximum reported carrier concentration at room temperature. The size of the circle indicates maximum achieved single-crystal wafer size. Normalized Baliga figure of merit calculated using measured quantities in state-of-art devices. Arrow near Ga$_2$O$_3$ and Al$_{0.85}$Ga$_{0.15}$N indicates that there are no known reports of room temperature p-type doping concentrations or measured hole mobility (and therefore p-type BFOM) for these materials. Most data obtained from 
    \cite{kaplar_generation-after-next_2017, tsao_ultrawide-bandgap_2018} and supplemented by material-specific data as listed in Table S1.} 
\end{figure*}

The semiconductors currently used for power applications sample a small portion of the materials that exist today. Their measured performance in low-frequency vertical devices as characterized by the well-known Baliga figure of merit \cite{baliga_semiconductors_1982} is shown in Fig.~\ref{fig:motivation} along with other indicators of performance and commercialization like the maximum realized carrier concentration at room temperature, thermal conductivity, and achievable wafer size. Silicon remains dominant in commercial power devices as a consequence of decades of research into economical and large-scale manufacturing processes. However, with a small critical electric field at breakdown \cite{kaplar_generation-after-next_2017}, silicon is hitting its theoretical performance limit within reasonable device size and power losses. 

Currently used alternatives to silicon include wide band gap ($1.5\,\,\mbox{eV}<\mbox{E}_g<3.4\,\,\mbox{eV}$) semiconductors such as GaN and SiC which offer significantly improved performance and smaller device size. The move toward wider band gap materials signifies a marked improvement in energy savings. For example, wide band gap metal-oxide-semiconductor field-effect-transistors cut heat losses by half as they increase efficiency from 96\% to 98\% compared to Si based devices \cite{tsao_ultrawide-bandgap_2018}. As shown in Fig.~\ref{fig:motivation}, the wide and ultra-wide band gap ($\mbox{E}_g\,>\,3.4$ eV) alternatives have Baliga figures of merit capable of addressing the demands of the future energy grid, especially \textit{n}-type. Unfortunately, considering only the semiconductor performance is not sufficient in determining commercial success. One must also consider other relevant factors like the ability to be doped (preferably both \textit{n} and \textit{p}-type), the ability to make high-quality crystals with large wafer size, and the ability to dissipate heat when operating at high power. For many of these materials, addressing these deficiencies is still a work in progress.

The handful of promising wide (WBG) and ultra-wide band gap (UWBG) power electronics in Fig.~\ref{fig:motivation} have resulted from decades of research into semiconductors by the microelectronics industry. With the exception of AlGaN alloys, all are binary compounds. Considering the many other binary as well as ternary and quaternary compounds that exist, one wonders whether there are other better performing semiconductors that have yet to be investigated.

Previous works that have proposed new WBG and UWBG semiconductors beyond those in Fig.~\ref{fig:motivation} often focus on one or two promising materials. Rutile GeO$_2$ is recently predicted to be an UWBG semiconductor with high electron mobility, ambipolar doping, and a Baliga FOM that surpasses current technologies \cite{bushick_electron_2020, chae_rutile_2019}. However, this performance has yet to be realized in real devices due to difficulty in thin film synthesis \cite{chae_epitaxial_2020, chae_toward_2021}.  Calculations have shown that metastable rocksalt ZnO is ambipolarly dopable unlike its ground state wurtzite version \cite{goyal_metastable_2018}, but experiments have yet to realize it. For these exploratory materials, single crystal growth, extrinsic doping, and device integration remain to be investigated.

\begin{figure*}
    \centering
    \includegraphics[width=\textwidth]{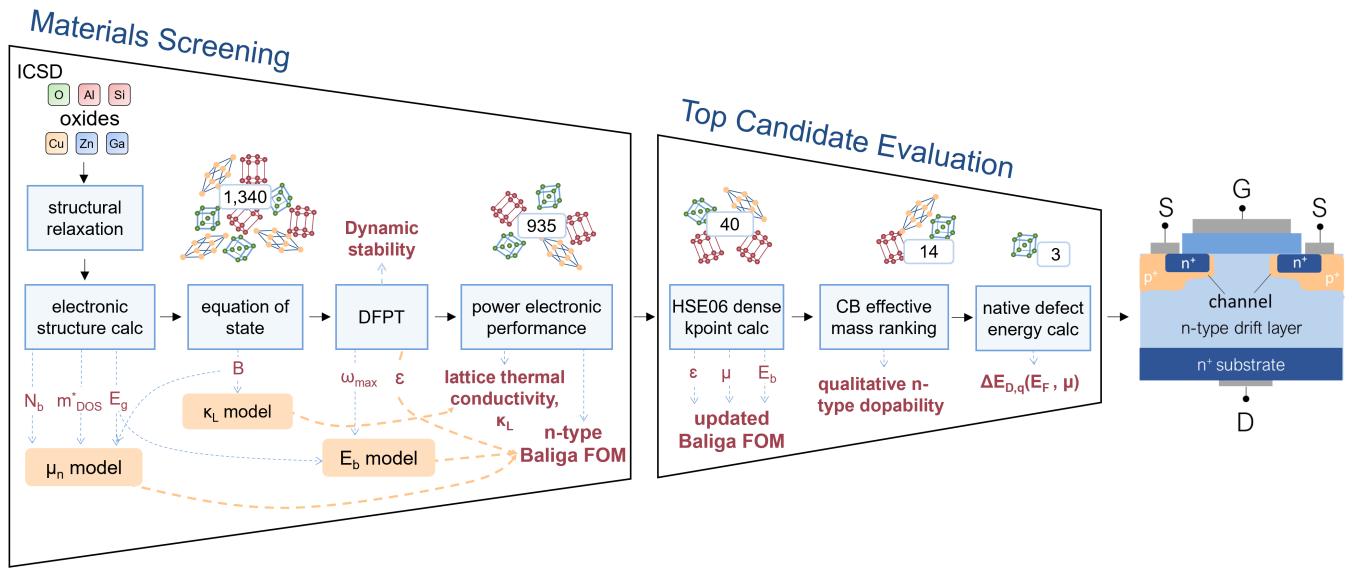}
    \caption{\label{fig:methods} Computational workflow to identify semiconductor oxides for power electronics. The first phase contains high-throughput GGA+U calculations to determine the Baliga FOM and lattice thermal conductivity ($\kappa_L$) and screen the materials. The second phase includes a more accurate calculation of the band structure using HSE06 hybrid functional for top candidates, a qualitative ranking of \textit{n}-type dopability, and detailed native defect formation energy calculations for the highest and lowest ranked candidates.}
\end{figure*}

Researchers have also begun to explore beyond traditional binary semiconductors, expanding to ternary UWBG compounds. Spinel Ga$_2$ZnO$_4$ has proven to be ambipolar dopable with a band gap similar to $\beta$-Ga$_2$O$_3$ and capable of bulk melt processing 
\cite{chikoidze_p-type_2020,galazka_ultra-wide_2019}. However, its low thermal conductivity prevents it from being a good candidate for high power applications. Similarly, inverse spinel Ga$_2$MgO$_4$ can be melt processed and doped \textit{n}-type, but experimental measurements of electron mobility are orders of magnitude lower than theoretically predicted performance and its lattice thermal conductivity is 
lower than that of $\beta$-Ga$_2$O$_3$ \cite{galazka_mgga2o4_2015}. LiGaO$_2$ is another UWBG material that is predicted to be \textit{n}-type dopable with Si or Ge, but experimental investigation of achievable carrier concentration is needed to prove its potential \cite{boonchun_first-principles_2019,dabsamut_first-principles_2020}.

In order to discover new candidates, it is advantageous to conduct a broad search rather than rely on intuition and historical bias. A more comprehensive computational search for novel power electronics semiconductors was recently performed by Gorai \textit{et al.} \cite{gorai_computational_2019}. They considered a select list of 863 sulfides, nitrides, carbides, silicides, borides, and oxides and evaluated Baliga figure of merit and thermal conductivity using high-throughput first-principles calculations. This screening identified a number of theoretically high-performing candidates, but did not systematically address whether these materials can be extrinsically doped to desired concentrations. The \textit{n}-type dopability, as determined by formation energy calculations of compensating native defects, was assessed for only a few select materials. But to fully understand which materials of the many candidates should be prioritized in experiments, one needs additional screening criteria. 

In this work we cast a wide net in the search for new power electronics among binary, ternary, and quaternary compounds. We use a combination of high-throughput screening using density functional theory (DFT) calculations and a more thorough analysis of top compounds to find new power electronic candidates from the Inorganic Crystal Structures Database (ICSD) \cite{hellenbrandt_inorganic_2004}. This study is focused on oxides which have the advantage of not oxidizing at high temperatures. Single crystals of oxides could also be easier to grow, as evidenced by $\beta$-Ga$_2$O$_3$ \cite{reese_how_2019}. We evaluate the performance of an extended list of oxides via the Baliga figure of merit (BFOM) and lattice thermal conductivity calculated using DFT calculated properties and semi-empirical transport models. We confirm the DFT results with a more accurate hybrid exchange-correlation functional (HSE06) on the select set of top DFT candidate materials. We also assess their \textit{n}-type dopability using a qualitative ranking based upon the conduction band effective masses. The dopability ranking is validated using modern defect theory and defect formation energy calculations. Lastly, we rely on literature to determine the synthesis history and potential for single crystal or thin film growth of these top candidates.

From this study, we have identified 3 novel families of ternary oxides, which have high theoretical performance based on the combination of their calculated Baliga FOM and thermal conductivity: (III)$_2$(IV)$_2$O$_7$ oxides with thortveitite and pyrochlore polymorphs;
(III)$_2$(II)O$_4$ and (II)$_2$(IV)O$_4$ oxides, most of them adopting the spinel crystal structure; and (III)BO$_3$ borates with the calcite structure. Many of the oxides from these families are also likely \textit{n}-type dopable according to our qualitative ranking. Through investigation of compositional stability and literature on experimental synthesis, we identify the most promising candidates within these families, many of which were previously grown as single crystals or are lattice matched to the common substrates for thin film growth. Taking into consideration the theoretical performance and potential for growth, we propose a select subset of 
these oxides, in particular  In$_2$Ge$_2$O$_7$, Mg$_2$GeO$_4$, and InBO$_3$, for further computational and experimental investigation for high power electronics. 

\section{Computational Methodology}
%


\subsection{Materials Screening}
%

The computational workflow used to determine the top performing oxides is shown in Fig.~\ref{fig:methods}. It is divided into two subsections: high-throughput materials screening, and higher-accuracy top candidate evaluation. 

\textbf{Performance metrics. }The materials screening section follows the methods outlined in Gorai \textit{et al.} \cite{gorai_computational_2019}. We use the well-known Baliga FOM for high-power, high voltage field effect transistors as the metric of performance \cite{baliga_semiconductors_1982}. The Baliga FOM takes into account the carrier mobility ($\mu$), dielectric constant ($\varepsilon$), and critical electric field at breakdown ($E_b$): 

\begin{equation}
    BFOM=\varepsilon\mu E_b^3.
    \label{eq:BFOM}
\end{equation}

The carrier mobility is important for conduction as it is a measure of how easily electrons or holes transport through a material. The dielectric constant and breakdown field set the maximum allowable applied voltage for a device with a given carrier concentration. Maximization of the Baliga FOM is equivalent to minimizing the on-state resistance of a conducting transistor in forward bias and thus indicates the maximum achievable efficiency for low-frequency devices. The BFOM does not, however, take into account the important role of thermal management in high power devices. The material must be able to quickly dissipate heat generated through power losses. Thus, we use the lattice thermal conductivity ($\kappa_L$) as an additional metric to score the performance of our materials.

\textbf{Search space. }Binary, ternary, and quaternary crystals containing oxygen as the sole anion were selected from the Inorganic Crystal Structures Database (ICSD) as the starting crystal structures \cite{hellenbrandt_inorganic_2004}. Cations consisted of metals and metalloids including likely non-magnetic transition metal elements (Sc, Y, La, Cu, Ti, Zr, Hf, Ta, W, Ag, Au, Zn, Cd, Hg) as an extension of the oxide data set generated in Gorai \textit{et al.} \cite{gorai_computational_2019}. Rare-earth elements and non-metals other than oxygen were excluded. Only stoichiometric compounds were considered. The search was also limited to compounds with unit cells of 50 atoms or fewer to make the task computationally tractable. There were 378 oxides repeated from Gorai \textit{et al.} and 959 additional transitional metal oxides for a total of about 1,340 oxides. 

\textbf{Calculation methods. }Starting structures were relaxed using the generalized gradient approximation (GGA) of Perdew-Burke-Ernzerhof (PBE) \cite{gga_pbe} in the projector augmented wave (PAW) formalism \cite{paw} as implemented in the Vienna Ab initio Simulation Package (VASP) \cite{vasp}. A plane wave cutoff of 340 eV was used and an on-site Hubbard \textit{U} correction (GGA+U) was applied for transition elements, following the methodology outlined in Ref.~\cite{Umethod,hubbardU}.

Once relaxed, electronic structure was calculated using a k-point grid density of $N \times n_{kpts} = 8000$, where $N$ is the number of atoms in the primitive cell and $n_{kpts}$ is the number of k-points. This dense k-point calculation served to predict the band degeneracy ($N_b$), density of states (DOS) effective mass for holes ($m^*_{DOS,vb}$) and electrons ($m^*_{DOS,cb}$), and electronic band gap ($E_g$). The DOS effective mass was calculated using a 100 meV window from the relevant band edges. The band effective masses ($m^*_b$) were then calculated as $m^*_{DOS} = N_b^{2/3}m^*_b$. At this point in the workflow, 272 compounds with $E_g=0$  were removed from subsequent calculations since high performing materials will need to have large band gaps. 

Next we performed total energy vs. volume calculations and fit the Birch-Murnaghan \cite{birch_finite_1947,murnaghan_finite_1937} equation of state to evaluate the bulk modulus ($B$). About 30 materials did not obey the Birch-Murnaghan equation of state, indicating a very shallow or even non-existent local minimum. These entries were removed from the dataset as they are likely unstable compounds. 

Next, a density functional perturbation theory (DFPT) calculation \cite{baroni_ab_1986,gajdos_linear_2006} was performed at the $\Gamma$ point only to evaluate the dielectric constant and phonon frequencies. The converged results were obtained using a stricter energy cutoff of 520 eV and k-point grid density of $N \times n_{kpts} = 1000$. The maximum phonon frequency ($\omega_{max}$) at $\Gamma$ and the directional average of the electronic $+$ ionic contributions to the dielectric tensor ($\varepsilon$) were extracted from the results. A total of 106 compounds had imaginary optical phonon modes, indicating that they are dynamically unstable. These were also removed from the dataset. The starting 1,340 structures were reduced to 935 at the end of the materials screening.

\textbf{Semi-empirical models. }The calculated values from the high-throughput DFT steps were used in previously-validated semi-empirical equations of the relevant material properties for the Baliga FOM. While direct calculations of the mobility, breakdown field, and lattice thermal conductivity are possible using DFT, the robust semi-empirical models make calculations of thousands of potential candidates more computationally tractable. The phonon-limited carrier mobilities, which are appropriate for high-temperature performance, were calculated using the previously-developed fitted model from Yan \textit{et al.} \cite{yan_material_2015}:

\begin{equation}
\mu=0.12B(m^*_b)^{-1.5}.
\label{eq:mobility}
\end{equation}

The critical breakdown field was calculated using a machine-learning fitted model developed from experimental band gaps and \textit{ab initio} calculations of breakdown field using electron-phonon coupling \cite{sun_intrinsic_2012, kim_organized_2016}:

\begin{equation}
    E_b=24.442 \,\, e^{0.315 \, \sqrt{\, E_g \, \omega_{max}}}.
\end{equation}

Finally, the lattice thermal conductivity was evaluated using a semi-empirical model previously validated in Miller \textit{et al.} \cite{miller_capturing_2017}:

\begin{equation}
    \kappa_L = A_1\frac{M\nu_s^y}{T\gamma^2V^zn^x} + A_2\frac{\nu_s}{V^z}\bigg{(}1-\frac{1}{n^{2/3}}\bigg{)}
\end{equation}
where M is the average atomic mass, $\nu_s$ is the speed of sound in m/s, $V$ is the volume per atom, $n$ is the number of atoms in the primitive cell, $\gamma$ is the Gruneisen parameter, $T$ is the temperature, and $A_1,A_2,x,y,z$ are fitted parameters. The speed of sound was calculated using $\nu_s\approx\sqrt{B/d}$, where $B$ is the bulk modulus $d$ is the density of the compound.

\begin{figure*}
    \includegraphics[width=\textwidth]{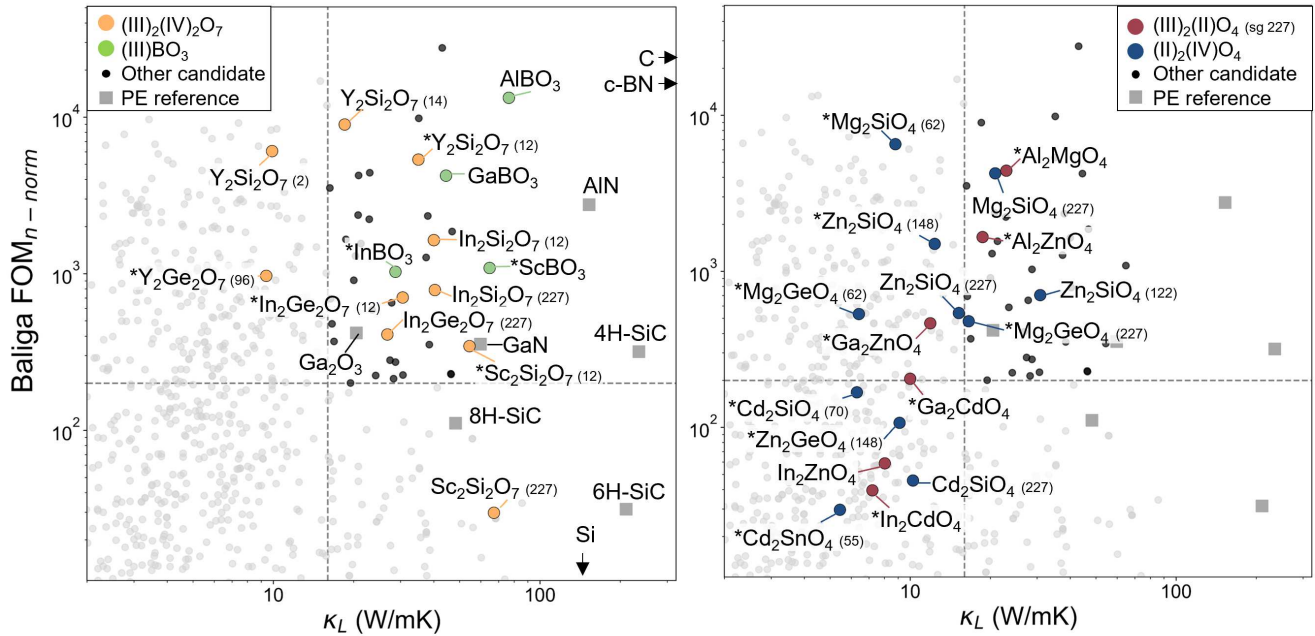}
    \caption{\label{fig:families} Baliga figure of merit for \textit{n}-type power electronic performance and lattice thermal conductivity of oxides calculated with GGA+U method. Highlighted top candidates (upper right quadrant) include members of one known: (III)$_2$(II)O$_4$ spinel (right), and three newly identified families of materials: (III)BO$_3$ calcite (left), (III)$_2$(IV)$_2$O$_7$ (left), and (II)$_2$(IV)O$_4$ (right) for power electronics. Numbers (eg: (227)) represent international space group number of the structure. Asterisks (*) denote ground state materials that lie on the convex hull.}
\end{figure*}

\subsection{Top Candidate Evaluation }
%
\textbf{Hybrid functional calculations.} Compounds with Baliga FOM and $\kappa_L$ above chosen cut-offs were down-selected as top candidates and subjected to more in-depth evaluation. First, a more accurate hybrid functional, HSE06 \cite{heyd_hybrid_2003,heyd_erratum_2006}), was used to re-relax and calculate the electronic structures. We used a standard mixing parameter of 0.25 and a screening parameter of 0.2 \AA$^{-1}$. While GGA+U is appropriate for high-throughput calculations, it is known to underestimate the band gaps \cite{marsman_hybrid_2008}. 
We chose to invest in higher-accuracy but computationally expensive hybrid calculations as additional verification of the BFOM ranking results. The equation of state calculations were not re-run with HSE06 since preliminary tests showed that bulk modulus calculated with the higher-accuracy method did not significantly impact Baliga figures of merit (see Table S4).

\textbf{\textit{N}-type dopability ranking.} Next, the top candidates were ranked according to a first-order approximation of their ability to be doped \textit{n}-type. \textit{N}-type dopability is typically assessed by calculating the formation energies of native compensating acceptor defects relative to the Fermi energy of the electron-doped material \cite{goyal_computational_2017}. Because calculations of defect formation energies are computationally expensive, this approach is not tractable for 40 materials. Instead, we used the HSE06-calculated conduction band (CB) effective mass ($m^*_{b,cb}$) as a proxy for \textit{n}-type dopability. 

The idea for using this method is inspired by Goyal \textit{et al.} who propose a complex set of governing material properties to determine dopability based upon a tight-binding model for binary ionic semiconductors \cite{goyal_dopability_2020}. One of these terms, $\bar\varepsilon_c - CBM$, describes the energy difference between the conduction band minimum ($CBM$) and the average electronic energy in the cation reference phase ($\bar\varepsilon_c$). If one considers $\bar\varepsilon_c$ to be close in energy to the middle of the conduction band, then the conduction band width must be large in order to maximize this term. A lighter effective mass (small $m^*_{b,cb}$) is correlated with more dispersive conduction bands and hence, larger band widths. Therefore, $m^*_{b,cb}$ can be used as a first order approximation of \textit{n}-type dopability, allowing us to rank the top candidates before performing computationally expensive defect formation calculations.

\textbf{Native defect calculations.}  Once ranked, the \textit{n}-type dopability of the best and worst candidates from the qualitative assessment were evaluated using detailed defect formation energy calculations of all relevant native point defects (vacancies, interstitials, antisites). We followed the standard supercell approach to calculate the formation energy of the defect $D$ in the charge state $q$ ($\Delta E_{D,q}$) as a function of Fermi energy (E$_F$) and at various chemical potentials ($\mu$) \cite{lany_accurate_2009, goyal_computational_2017}. We used an 88-atom supercell with 2x2x2 k-point mesh for In$_2$Ge$_2$O$_7$, a 297-atom supercell with gamma only mesh for CaZrSi$_2$O$_7$, and a 360-atom supercell with gamma only mesh for AlScO$_3$. The defect supercells were relaxed with DFT-GGA (+U, U=3.0 for Zr and Sc) using a planewave cutoff energy of 340 eV and a force convergence criterion of 5 meV/\AA. Further, self-consistent GW calculations with fixed wavefunctions \cite{shishkin_self-consistent_2007} were performed to correct the band gap error of GGA+U. The chemical potentials were calculated using the fitted elemental-phase reference energy approach \cite{stevanovic_correcting_2012}. Static dielectric constants of 21.85, 10.68, and 18.96 were calculated using GGA+U and were used for the image-charge correction \cite{goyal_computational_2017} for In$_{2}$Ge$_{2}$O$_{7}$, CaZrSi$_{2}$O$_{7}$ and AlScO$_3$, respectively.

\textbf{Synthesis evaluation. }Lastly we studied the literature for indication of previous synthesis of these materials. This is especially important for those compounds which are not ground state materials as they may decompose to competing phases. In these cases we looked for evidence of synthesis and ambient condition stability through high pressure and quenching techniques. In addition, we noted any published demonstration of single crystal growth as this could allow bulk growth of native substrates. For select top candidates, we also gauged the possibility of epitaxial growth of thin film heterostructures by comparing the lattice constants of the candidate compound and well-known substrates. Understanding material stability and synthesis is key in prioritizing which materials to investigate further with experimental efforts.

\begin{figure*}
    \includegraphics[width=0.7\textwidth]{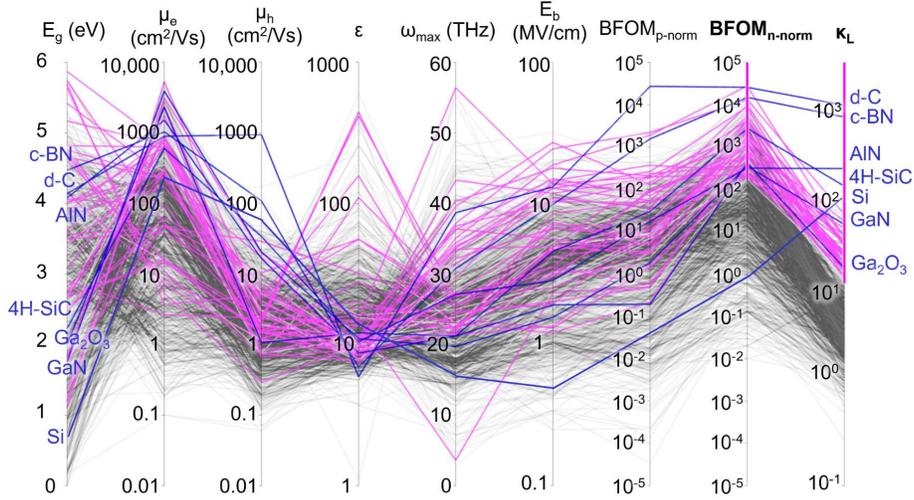}
    \caption{ \label{fig:GGA_results} Parallel coordinate plot of calculated electronic properties from DFT GGA+U for all oxide candidates and the resultant \textit{n}-type and \textit{p}-type Baliga figures of merit. Blue lines represent the reference power electronic materials. Magenta lines represent all those candidates which fulfill the cutoff criteria for top candidates: $\kappa_{L}$ = 16 W/mK and BFOM$_{\mbox{n-norm}}=$ 200.}
\end{figure*}

\section{Performance Metric Results}
%

\subsection{Materials Screening: DFT-GGA Calculations} Using the structure selection criteria, we start with a total of around 1,340 oxide crystals from the ICSD. After completing the high-throughput DFT workflow, there are 925 remaining structures which are dynamically stable and have finite band gaps. The thermal conductivity vs. Baliga FOM for all calculated structures is plotted in Fig.~\ref{fig:families}. The axes limits have been chosen to magnify the results of the best performers. Baliga figure of merit values have been normalized to Si \textit{n}-type (Si BFOM$_{\mbox{n-norm}}$=1). Commercially-used (GaN, SiC, Si) and previously proposed ($\beta$-Ga$_2$O$_3$, BN, AlN, diamond) power electronic semiconductors are indicated with the squares as a reference. The reference materials follow the expected trends. Si has the lowest BFOM (below limits of the y-axis), $\beta$-Ga$_2$O$_3$ has the lowest $\kappa_L$, and diamond and cubic BN perform an order of magnitude above the rest of the reference materials in both measures. 

When one considers the important commercialization factors beyond figure of merit, it becomes clearer that new alternatives to common power electronic materials would be helpful. Despite the theoretical potential for high performance, diamond's application in power electronics is limited by its small wafer size, extreme fabrication costs, high dislocation densities, and lack of high-quality substrates for epitaxial growth \cite{tsao_ultrawide-bandgap_2018}. AlN is limited by the absence of heavily doped bulk substrates and only 10$^{15}$ cm$^{-3}$ achievable electron concentration through Si doping \cite{tsao_ultrawide-bandgap_2018, wang_2021}. Cubic BN suffer from similar substrate challenges \cite{tsao_ultrawide-bandgap_2018}. Low Al content AlGaN has the potential for growth on GaN substrates, albeit with some concern for interface strain, but does not provide significant performance advantages over other WBG materials \cite{tsao_ultrawide-bandgap_2018,allerman_2016}. High Al content Al$_{0.85}$Ga$_{0.15}$N is more appropriate for high power vertical devices, but has limited wafer size and is not \textit{p}-type conductive due to high activation energy of Mg dopants \cite{pearton_gan_2012,sarkar_2018,tsao_ultrawide-bandgap_2018}. Even GaN and 4H-SiC, which are the state-of-the-art choices for high frequency and high power electronics, have their challenges. SiC is still relatively costly to produce with low defect densities \cite{huang_power_2017,pearton_perspective_2018}. GaN can be grown on 200 mm Si substrates but lacks decent sized vertical devices grown on a native substrate which are critical for high power applications \cite{pearton_perspective_2018}.

Our search results identify plenty of potential materials that could complement and compete with current PE semiconductors. To downselect, we set the top candidate cutoff values at 4H-SiC for BFOM$_{\mbox{n-norm}}$ and $\beta$-Ga$_2$O$_3$ for $\kappa_L$ plus a margin. Using this criteria, we identify 40 oxides with BFOM$_{\mbox{n-norm}}>$ 200 and $\kappa_L>$ 16 W/mK which are plotted in the top-right of Fig.~\ref{fig:families}. A few interesting families of materials, which are discussed in detail later, are labeled as well. Of these top candidates, 85\% are ternaries, despite them making up 55\% of the dataset. Given that most incumbent materials are binary compounds, the discovery of so many promising ternary compounds points to the impact that a broader computational search can have on next-generation power electronics.

Diving deeper into the results, let us first inspect the relative influence of each material property on the top candidates. Fig.~\ref{fig:GGA_results} shows the parallel coordinate plot of the GGA-calculated values for the governing material properties that make up the \textit{n}-type and \textit{p}-type Baliga FOM of all screened materials (see also Supplemental Data). The top 40 candidates are highlighted in magenta and the reference materials in blue. As a whole, most candidates exhibit better \textit{n}-type figure of merit than \textit{p}-type, due to lighter electron effective masses than hole effective masses. This fits with expectations given most known metal oxide semiconductors show electron conduction \cite{he_2_2020, goyal_dopability_2020} and commercial power devices utilize \textit{n}-type Si, SiC, and GaN over \textit{p}-type. This provides justification for limiting our search to \textit{n}-type candidates. 

The normalized \textit{n}-type BFOM spans seven orders of magnitude, but the top candidates are all within $10^2$ and $10^5$ BFOM$_{\mbox{n-norm}}$. Despite the Baliga FOM's heavy dependence on the breakdown field, $E_b$ is not the only driver in performance, as indicated by top candidates with breakdown fields that fall in the bottom half of the dataset. This is further evidenced by such a large spread in the top candidates' band gaps which is a prominent property used in the calculation of the breakdown field. While the expected ultrawide band gap candidates do appear, there are also compounds, like rutile GeO$_2$ and HgTiO$_3$, with band gaps smaller than WBG 4H-SiC and GaN. Ultra high electron mobilities and/or dielectric constants make these compounds top power electronic candidates. This suggests it may be necessary to rethink the recent strong emphasis on ultra-wide band gaps as the leading criteria for new candidates, as these compounds would have been missed. 

Lastly, we notice that, in general, the oxides have lower lattice thermal conductivity than the reference carbides and nitrides which have lighter mass anions and greater bulk moduli. This is a trade-off we must consider when for elevated temperature applications where oxide chemistries are preferred.

\subsection{Top Candidate Evaluation: HSE06 Calculations }To confirm the best candidates, the band structure properties of the top 40 candidates and reference materials were re-calculated using HSE06 hybrid functional. Electronic properties from HSE06 in comparison to GGA+U calculations can be found in supplemental Tables S2 and S3. Materials with the highest GGA+U electron mobilities showed an increase in the band effective mass with HSE06 and thus, a decrease in electron mobility. But all mobility changes were less than an order of magnitude from the original value. The band gap increased by up to 2.6 eV in all of the materials. Given the importance of band gap in the calculation of Baliga figure of merit, this increase resulted in a higher HSE06 Baliga figure of merit for all but two of our materials. Despite smaller BFOM$_{\mbox{n-norm}}$ for GeHfO$_4$ and CuTaO$_3$, these and all of the other leading candidates confirm their membership as predicted top performers with HSE06 calculations.

\section{\textit{N}-type Dopability Assessment}

\subsection{Qualitative Ranking}
%
\begin{figure}
    \includegraphics[width=0.93\columnwidth]{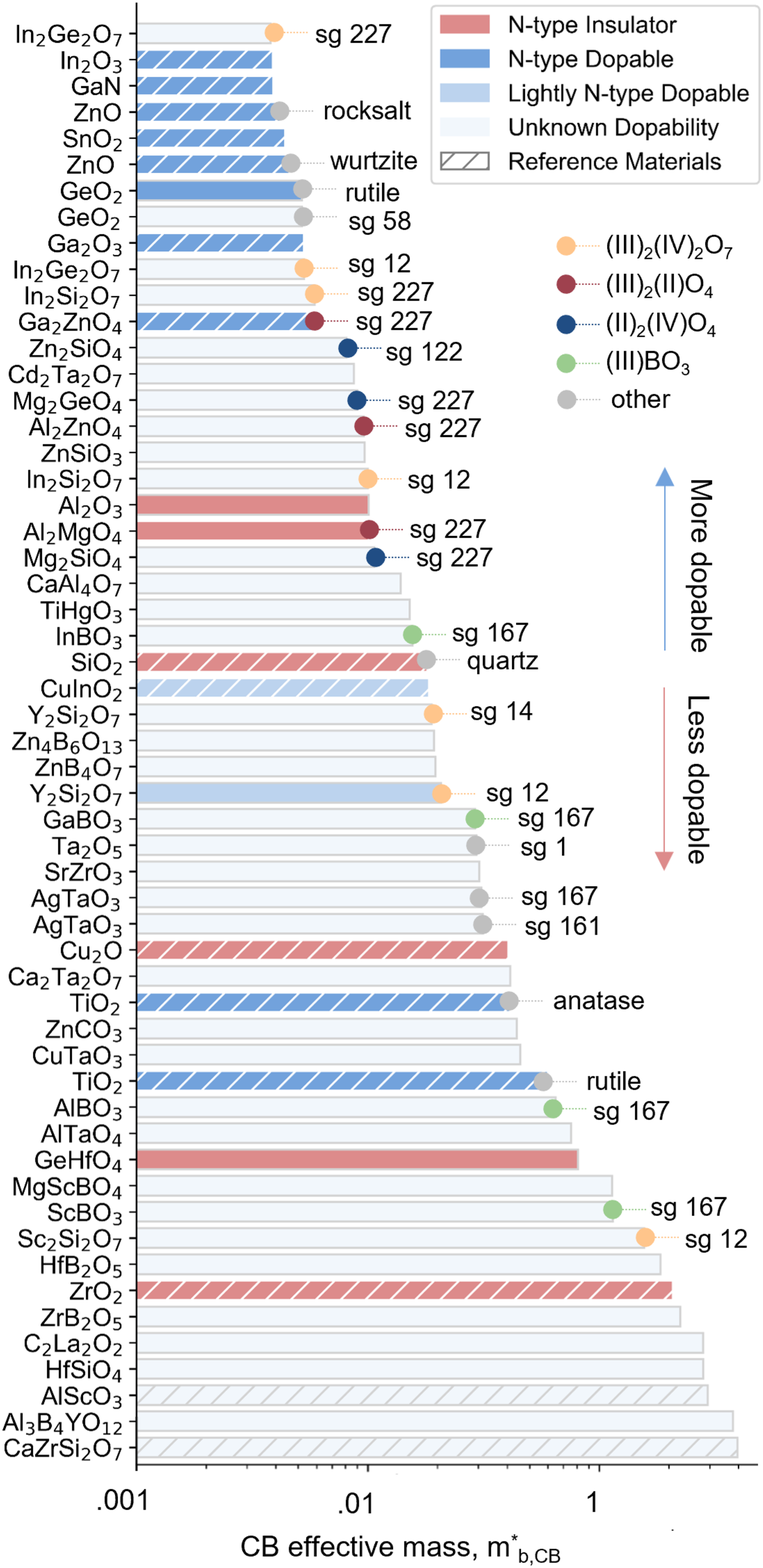}
    \caption{\label{fig:dopability} Qualitative assessment of \textit{n}-type dopability by HSE06 calculation of conduction band effective mass ($m^*_{b,CB}$). Smaller $m^*_{b,CB}$ indicates a more dopable compound. Top 40 oxide candidates shown along with known \textit{n}-type insulators and \textit{n}-type dopable materials for comparison. Families of materials and the compounds' specific space groups (sg) are identified with colored circles. HSE06 calculated values for $m^*_{b,CB}$ found in Tables S2 and S3. }
\end{figure}

Any material used as a conducting channel in power electronics must be able to be extrinsically doped to desired concentrations. The 40 top candidates are ranked according to their HSE06-calculated conduction band effective mass ($m^*_{b,cb}$) as shown in Fig.~\ref{fig:dopability}, where smaller indicates more likely \textit{n}-type dopable. The ranking includes known \textit{n}-type conducting and \textit{n}-type insulating materials to gauge the method effectiveness. \textit{N}-type dopable reference materials are defined as having 10$^{18}-10^{21}$ cm$^{-3}$ electron carrier concentrations. Lightly \textit{n}-type dopable materials are those which can achieve 10$^{16}-10^{18}$ cm$^{-3}$ carrier concentrations. Many of the reference carrier concentrations are found in the supplementary material in Goyal \textit{et al.} \cite{goyal_dopability_2020}. 

At the top of Fig.~\ref{fig:dopability} appear known \textit{n}-type dopable semiconductors. In$_2$O$_3$ can be \textit{n}-type doped with Sn to degenerate concentrations \cite{groth_untersuchungen_1966}. GaN, SnO$_2$, $\beta$-Ga$_2$O$_3$ and wurtzite ZnO are common semiconductors which can be \textit{n}-type doped up to $10^{20}$ cm$^{-3}$. The III-II spinel, Ga$_2$ZnO$_4$, is has been moderately \textit{n}-type doped as well \cite{goyal_dopability_2020}. Recent computational studies of rocksalt ZnO determined that it is likely \textit{n}-type dopable  \cite{goyal_metastable_2018}. Hybrid calculations of rutile GeO$_2$ predict ambipolar dopability and strong power electronic performance \cite{chae_rutile_2019}. This corroborates the findings of our ranking. Unfortunately, none of the known \textit{n}-type reference materials mentioned in Fig.~\ref{fig:dopability}, besides GaN and $\beta$-Ga$_2$O$_3$, have large enough Baliga FOM and $\kappa_L$ to be considered a top power electronics candidate (see also Table S3). 

In the bottom half of Fig.~\ref{fig:dopability} are those materials with the heaviest conduction band effective masses including known \textit{n}-type insulators ZrO$_2$ and Cu$_2$O \cite{goyal_dopability_2020}. Defect formation energy calculations have shown that one of our candidates, GeHfO$_4$, is also likely insulating \cite{gorai_computational_2019}. 

In the middle of the figure, the trend is less clear. There are known insulators like quartz SiO$_2$, spinel Al$_2$MgO$_4$, and corundum Al$_2$O$_3$, as well as lightly \textit{n}-type dopable materials like CuInO$_2$ and the recently calculated Y$_2$Si$_2$O$_7$ \cite{gorai_computational_2019, goyal_dopability_2020}. Both anatase and rutile TiO$_2$ are known to be \textit{n}-type dopable yet their conduction band effective masses are quite large. This is due to the appearance of Ti \textit{d}-orbitals at the bottom of the conduction band. While this qualitative ranking is not perfect, it's clear that the smaller the $m^*_{cb}$ the more likely the material can be doped \textit{n}-type without compensation by acceptor defects if a suitable extrinsic dopant is found.

Looking at the top candidates more closely, we see common chemistries and stoichiometries emerge, creating "families" of materials. The highest-ranked compound in Fig.~\ref{fig:dopability} is In$_2$Ge$_2$O$_7$ pyrochlore (space group number(s.g.) 227). It's ground state polymorph (s.g. 12) and silicate versions appear in the top half of Fig.~\ref{fig:dopability} as well. These are part of a family of (III)$_2$(IV)$_2$O$_7$ compounds and are indicated by yellow dots in Figs. \ref{fig:families} and \ref{fig:dopability}. Next we see the appearance of (III)$_2$(II)$_2$O$_4$ (red dot) and (II)$_2$(IV)$_2$O$_4$ (blue dot) compounds, many of which have a normal spinel crystal structure. The (III)$_2$(II)$_2$O$_4$ spinels are a known category of WBG materials that have already been explored for use in electronics while the (II)$_2$(IV)$_2$O$_4$ compounds have not. Lastly, there are four (III)BO$_3$ borates with calcite crystal structure (s.g. 167). These do not rank as highly in the \textit{n}-type dopability analysis as the other families but have incredible Baliga figures of merit as shown by the green dots in Fig.~\ref{fig:families}.  

Taking InBO$_3$ as the cutoff for qualitative ranking of \textit{n}-type dopability, we are able to downselect from 40 to 14 candidates which were not previously considered for power electronics. The HSE06 calculated electronic properties of these 14 materials are shown in Table \ref{tab:HSE-results}. Many of these 14 most promising candidates are also members of the identified families.

\begin{table*}
\caption{\label{tab:HSE-results} Computed electronic structure properties from HSE06 calculations of the top 14 previously unexplored \textit{n}-type power electronic candidates. These candidates pass the \textit{n}-type BFOM and $\kappa_L$ criteria and are in the top half of the qualitative \textit{n}-type dopability ranking. Shown here are the international space group number (s.g.\#), band gap ($E_g$), band effective mass for holes ($m^*_{b,vb}$) and electrons ($m^*_{b,cb}$), mobility of holes ($\mu_p$) and electrons ($\mu_n$), breakdown field ($E_b$), and \textit{p}-type and \textit{n}-type Baliga figure of merit. The Baliga FOM has been normalized to the calculated \textit{n}-type Silicon HSE06 BFOM. Candidates are listed in alphabetical order. For more calculated compounds see Table S2.}
\fontsize{9}{12}\selectfont
\begin{ruledtabular}
\begin{tabular}{ccp{1.2cm}ccccccc}
           &   & $E_g$ & $m^*_{b,vb}$ & $m^*_{b,cb}$ & $\mu_p$ & $\mu_n$ & $E_b$    &   &  \\ 
Compound             & s.g.\#  & (eV) & ($m_{e}$) & ($m_{e}$) & (cm$^2$/Vs) & (cm$^2$/Vs) & (MV/cm)    & BFOM$_{\mbox{p-norm}}$  & BFOM$_{\mbox{n-norm}}$ \\ \hline
Al$_2$ZnO$_4$     & 227 & 6.0 & 2.88        & 0.10        & 4.6     & 758.3   & 9.5 & 21            & 3590             \\
CaAl$_4$O$_7$     & 15  & 5.8 & 3.69        & 0.14        & 1.9     & 258.8   & 13.0 & 24             & 3290             \\
Cd$_2$Ta$_2$O$_7$ & 227 & 3.7 & 6.06        & 0.09        & 1.2     & 715.6   & 5.6 & 14            & 8560             \\
GeO$_2$           & 58  & 3.9 & 0.98        & 0.05        & 24.0    & 1940.2  & 5.0 & 27             & 2210             \\
HgTiO$_3$         & 161 & 2.8 & 2.36        & 0.15        & 5.8     & 350.9   & 2.8 & 12             & 781             \\
In$_2$Ge$_2$O$_7$ & 227 & 3.2 & 6.08        & 0.04        & 1.5     & 3043.5  & 3.7 & 0.9             & 1910             \\
In$_2$Ge$_2$O$_7$ & 12  & 4.0 & 5.63        & 0.05        & 1.1     & 1252.8  & 6.7 & 3             & 3520             \\
In$_2$Si$_2$O$_7$ & 227 & 3.8 & 3.85        & 0.06        & 3.6     & 1909.8  & 5.1 & 5             & 2930             \\
In$_2$Si$_2$O$_7$ & 12  & 4.8 & 3.68        & 0.10        & 2.5     & 557.7   & 13.3 & 33             & 7490             \\
InBO$_3$          & 167 & 4.8 & 2.45        & 0.16        & 4.5     & 278.6   & 15.7 & 104            & 6480             \\
Mg$_2$GeO$_4$     & 227 & 5.3 & 2.92        & 0.09        & 3.8     & 696.8   & 7.1 & 7             & 1350             \\
Mg$_2$SiO$_4$     & 227 & 7.0 & 4.31        & 0.11        & 2.3     & 599.0   & 15.0 & 37             & 9670             \\
Zn$_2$SiO$_4$     & 122 & 4.7 & 7.63        & 0.08        & 0.8     & 689.1   & 8.4 & 2            & 2350             \\
ZnSiO$_3$         & 148 & 5.9 & 2.80        & 0.10        & 4.4     & 685.4   & 9.3 & 29             & 4620            
\end{tabular}
\end{ruledtabular}
\end{table*}

%
\subsection{Defect Formation Energy Calculations}

\begin{figure}
\includegraphics[width=0.93\columnwidth]{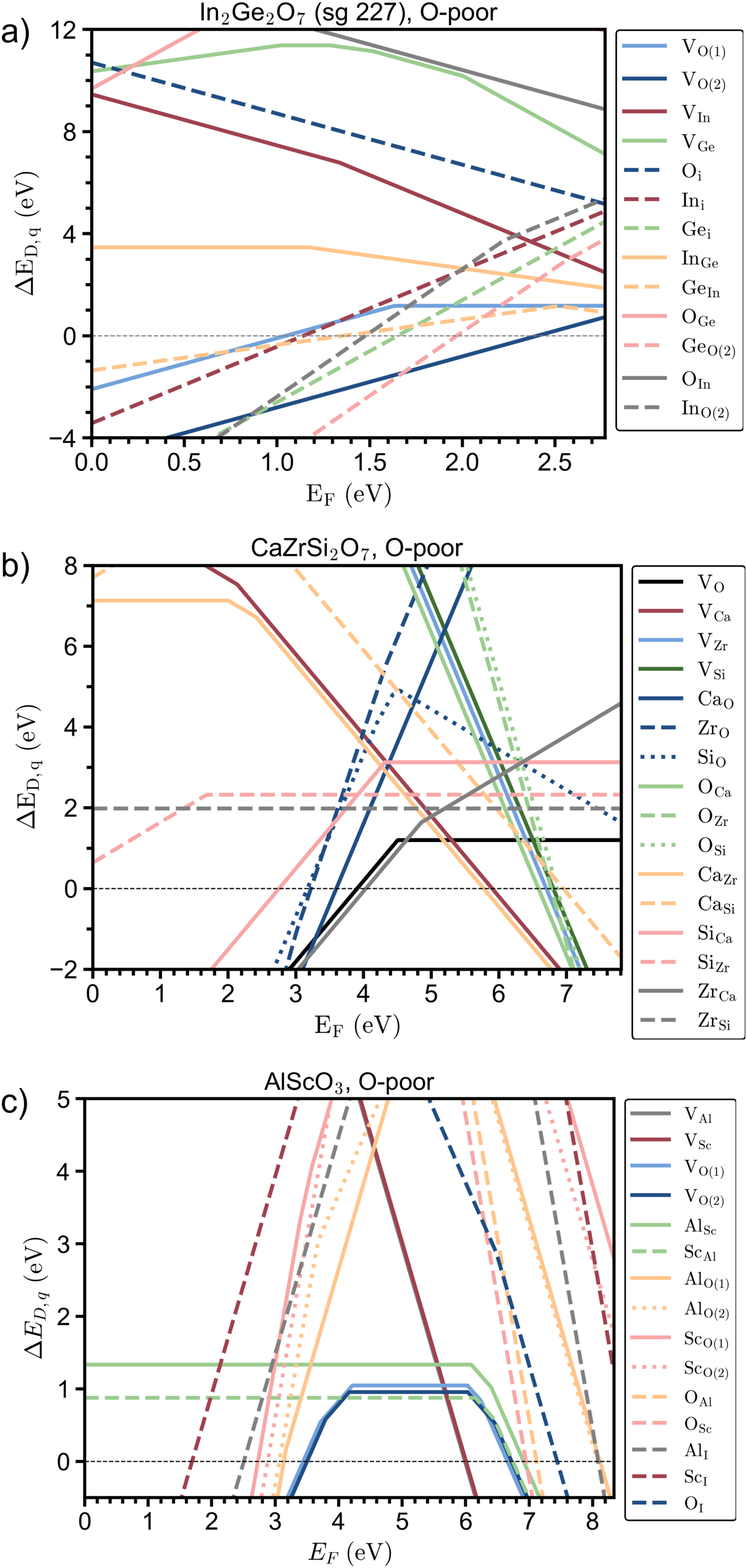}
 \caption{\label{fig:defect} Formation energies of native vacancy, interstitial, and anti-site defects calculated with GGA+U with GW band gap correction. The Fermi energy is referenced from the valence band maximum and plotted to the conduction band minimum. Subscripts (1) and (2) indicate different Wycoff sites. (a) Pyrochlore In$_2$Ge$_2$O$_7$ is predicted to be \textit{n}-type dopable under Ge-rich conditions. CaZrSi$_2$O$_7$ (b) and AlScO$_3$ (c) are predicted to be \textit{n}-type insulating even in favorable O-poor conditions.}
\end{figure}

To add more confidence to the qualitative \textit{n}-type dopability ranking, defect formation energy calculations were performed on three structures: highest-ranking In$_2$Ge$_2$O$_7$ (s.g. 227), and low-ranking AlScO$_3$ and CaZrSi$_2$O$_7$. Native acceptor defects with low formation energy can accept electrons and, if present in large numbers, will compensate any dopant electrons, limiting the achievable free carrier concentration. Ideally, one would like acceptor defect formation energies to become negative at a Fermi level that is inside the conduction band as this will allow dopants to increase the net electron concentration without spontaneous formation of compensating defects \cite{zunger_practical_2003}. Fig.~\ref{fig:defect} shows the formation energies calculated using the GGA+U method for all possible vacancy (V$\mathrm{_{A}}$), interstitial (A$\mathrm{_{i}}$), and anti-site (A$\mathrm{_{B}}$) point defects for each element (A) in In$_2$Ge$_2$O$_7$ (s.g. 227), AlScO$_3$, and CaZrSi$_2$O$_7$. The x-axis is the Fermi energy in eV referenced from valence band maximum to conduction band minimum and represents the corrected band gap as calculated using the GW method.

The germanium-rich condition is the most favorable environment for \textit{n}-type doping in pyrochlore In$_2$Ge$_2$O$_7$ as shown in Fig.~\ref{fig:defect}(a). Under these conditions, the lowest-energy compensating defect, Ge$\mathrm{_{In}}$, will spontaneously form at a Fermi energy that is $\approx1$ eV above the conduction band. This means that there is likely room for increasing the Fermi energy and electron concentration through extrinsic donor dopants if a suitable one can be found. 

The opposite case is shown in Fig.~\ref{fig:defect}(b-c). CaZrSi$_2$O$_7$ and AlScO$_3$ are \textit{n}-type insulating even in oxygen-poor conditions where \textit{n}-type doping is most favorable. If donor dopants are added to CaZrSi$_2$O$_7$, the additional electrons will be compensated by native acceptor defects (i.e. Ca$\mathrm{_{Zr}}$, V$\mathrm{_{Ca}}$) pinning the Fermi energy $\approx$1.2 eV below the conduction band minimum. For AlScO$_3$, the cation vacancies sit $>$2 eV below the conduction band minimum, preventing \textit{n}-type dopants from increasing the electron concentration in this material as well. These defect formation energy results confirm the predictions of the qualitative dopability ranking and further support pyrochlore In$_2$Ge$_2$O$_7$ as one of the most promising among the top power electronics candidates from our search.

\section{Discussion: Synthesis of Top Candidates}

Another important factor for commercialization of semiconductors is the ability to synthesize as single crystals and/or thin films. Understanding whether and how the candidates have been historically grown will help researchers prioritize experimental investigations among those materials with promising predicted performance and dopability. Specifically, materials which can easily be grown as single crystals (\textit{e.g.} Si, $\beta$-Ga$_2$O$_3$) allow melt-based bulk processing of substrate wafers which are then used to grow homoepitaxial thin films of the same material \cite{reese_how_2019}. In the epitaxial growth method, new crystal layers are formed with desired orientation governed by the underlying substrate lattice. Doped native substrates are required for fabrication of vertical devices like those made with SiC, and thus, are preferred for high power applications. Alternatively, one can grow heteroepitaxially on non-native substrates which are closely lattice-matched to the semiconductor. Heterostructures are used in the creation of lateral devices (\textit{e.g.} GaN on Si) which are less suited to high power application, but they are also useful in the early stages of lab testing for making high-quality measurements of a material's electrical properties. Thus, we are interested in assessing both the single crystal growth and availability of closely matched non-native substrates for our top candidates.
To address this synthesis and device integration potential, we considered the growth of the aforementioned novel families of materials: (III)$_2$(IV)$_2$O$_7$, (III/II)$_2$(II/IV)$_2$O$_4$, and (III)BO$_3$. Considering synthesis for material groups is useful since many compounds in the same family will share growth recipes and have similar material behaviors. 

We analyze the available literature for each family to answer a few pertinent questions: 1) Can the material be synthesized and remain stable at ambient conditions? 2) Has the material been grown as a single crystal? 3) Is there demonstrated electrical conductivity or successful extrinsic doping? 4) Could the material be grown heteroepitaxially with a suitable substrate?

\subsection{(III)$_2$(IV)$_2$O$_7$ family}

The first family among top candidates is a group of 2-2-7 stoichiometry oxides of the form (III)$_2$(IV)$_2$O$_7$ where the group III elements include In, Sc, or Y, and group IV elements are Si or Ge. The In and Sc versions of this chemistry appear in two polymorphs: a ground state thortveitite structure (s.g. 12) and a higher pressure pyrochlore structure (s.g. 227). With its larger atomic radii, the yttrium versions have a different set of polymorphs. Y$_2$Si$_2$O$_7$ appears as thortveitite ground state and several higher temperature monoclinic and orthorhombic polymorphs \cite{ito_synthesis_1968}.  Yttrium germanate's ground state is space group 96. According to literature, a pyrochlore Y$_2$Ge$_2$O$_7$ \cite{shannon-sleight_Inorg_1968,redhammer_yttrium_2007} and several Sc$_2$Ge$_2$O$_7$ polymorphs have been synthesized in experiments \cite{shannon-sleight_Inorg_1968,li_-situ_2020,shannon_prewitt_synthesis_1970},  but none of these chemistries are included in the ICSD and are therefore not in this study. Of the ten ICSD structures in the (III)$_2$(IV)$_2$O$_7$ family, seven are top candidates for power electronics (see Fig.~\ref{fig:families}). In addition, the In versions of both thortveitite and pyrochlore structure rank within the upper half of our \textit{n}-type dopability assessment (see Fig.~\ref{fig:dopability}). Native defect formation energy calculations completed in this work for pyrochlore In$_2$Ge$_2$O$_7$ further indicate that materials in this family may allow \textit{n}-type doping.

Powders of the thortveitites have been synthesized using conventional solid state reactions at elevated temperatures (1000-1500$^\circ$C) \cite{li_comparative_2015, li_-situ_2020,shannon_prewitt_synthesis_1970,shannon-sleight_Inorg_1968,reid_high-pressure_1977}. The pyrochlores can be stabilized in powder form using high pressure: 5.3 GPa for pyrogermanates and 12 GPa for pyrosilicates \cite{reid_high-pressure_1977,li_comparative_2015,li_-situ_2020}. The thortveitite In$_2$Si$_2$O$_7$ has also been grown as single crystals of a few mm$^2$ and 1 mm thick using a flux method and Li$_2$Mo$_2$O$_7$ solvent \cite{messous_indium_1995}. Previous research into these materials focuses on fluoresence centers, catalysts, oxide fuel cells, and exotic behaviors like superconductivity, but not power electronics \cite{li_-situ_2020, li_comparative_2015}. 

While our calculations show that In$_2$Ge$_2$O$_7$ pyrochlore is likely n-type dopable, its metastable state and high-pressure synthesis may make bulk growth difficult. In contrast, the thortveitite structure has potential for better high-power performance (see Table \ref{tab:HSE-results}), ranks in the top half of Fig.~\ref{fig:dopability}, and could be stably grown in bulk as a single crystal. This makes the thortveitite In$_2$Ge$_2$O$_7$ and In$_2$Si$_2$O$_7$ attractive top candidates for power electronic devices. 

\begin{figure}
    \includegraphics[width=\columnwidth]{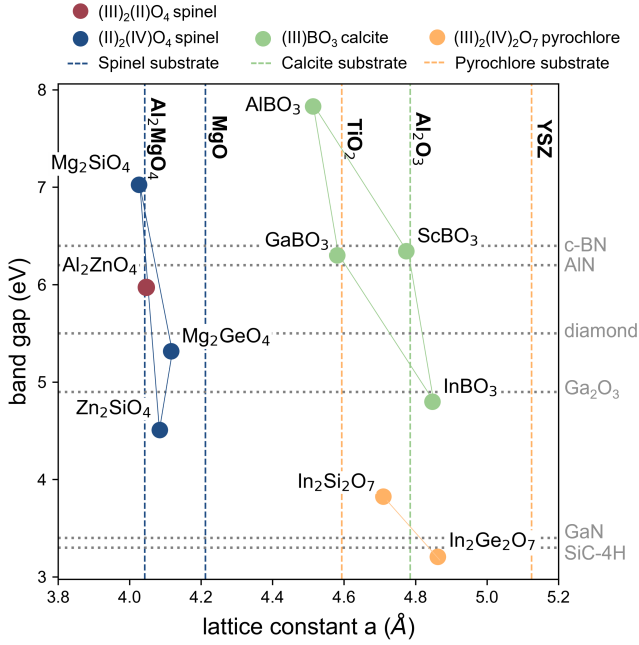}
    \caption{\label{fig:lattice_map} Power electronic candidates from the three promising families: (III)$_2$(IV)$_2$O$_7$ pyrochlore, (III/II)$_2$(II/IV)O$_4$ normal spinels, and (III)BO$_3$ calcite. HSE06 calculated band gaps of candidates shown in comparison to experimental band gaps of known power electronic semiconductors. HSE06 calculated "a" lattice constant of candidates shown in comparison to experimental lattice constants of potential substrates (color matched) for heteroepitaxial growth. Half lattice constant shown for Al$_2$MgO$_4$ substrate, (II)$_2$(IV)O$_4$ spinels, and (III)$_2$(IV)$_2$O$_7$ pyrochlores.}
\end{figure}

\begin{figure}
    \includegraphics[width=0.9\columnwidth]{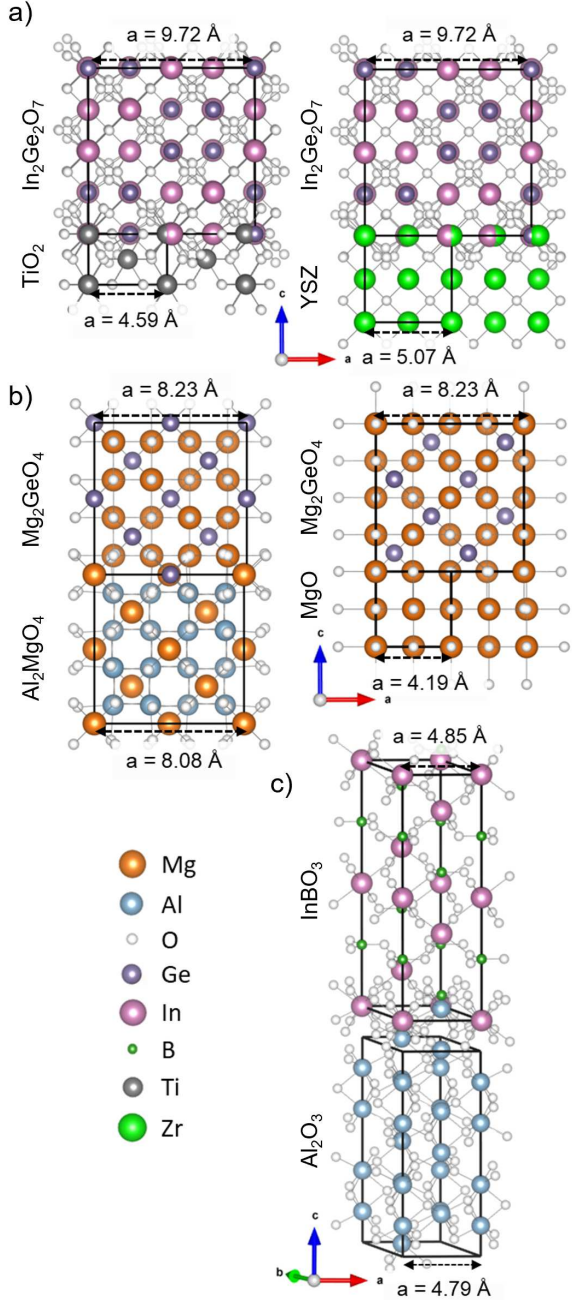}
    \caption{\label{fig:lattice_vesta} Lattice matching schematic of representatives from the three promising families with potential substrates: a) In$_2$Ge$_2$O$_7$ pyrochlore (s.g. 227) on TiO$_2$ (s.g. 136) and Yttria-stabilized ZrO$_2$ (s.g. 225), b) Mg$_2$GeO$_4$ spinel (s.g. 227) on Al$_2$MgO$_4$ (s.g. 227) and MgO (s.g. 225), and c) InBO$_3$ (s.g. 167) on Al$_2$O$_3$ (s.g. 167).}
\end{figure}

In order to synthesize the pyrochlore versions, we will need to consider heteroepitaxial growth of thin films. Fig.~\ref{fig:lattice_map} shows the HSE06 relaxed lattice constant (a) of the pyrochlores along with a few common substrates: TiO$_2$ and cubic yttria-stabilized zirconia (YSZ). The lattice mismatch between semiconductor and these two substrates is between 2\% and 6\% which is within the acceptable tolerance. Fig.~\ref{fig:lattice_vesta} shows a schematic of In$_2$Ge$_2$O$_7$ crystal on both substrates, displaying the close alignment of the cations that is necessary for successful epitaxial growth. Given their indication of \textit{n}-type dopability according to calculations and potential for heteroepitaxy, the pyrochlore 2-2-7 materials also have potential for power applications along with their ground state versions. 

%
%

\subsection{(III/II)$_2$(II/IV)O$_4$ family}

%

The next family of materials is 2-1-4 oxides. Within this stoichiometry, there are well-known spinels (s.g. 227) of chemistry (III)$_2$(II)O$_4$  where group III elements are Al, Ga, In and group II elements are Mg, Zn, Cd. There are also (II)$_2$(IV)O$_4$ compounds where group II elements are Mg, Zn, Cd, and group IV elements are Si, Ge, Sn. Several of the (II)$_2$(IV)O$_4$ compounds also take the spinel structure. All of the spinels calculated here are of the "normal" structure where the divalent B atoms of A$_2$BO$_4$ sit on the 8 tetrahedral sites and the trivalent A atoms sit on the 16 octahedral sites. There also exist "inverse" spinels where a 100\% inverted structure has all B atoms and half of A atoms occupy octahedral sites and the rest of A atoms occupy tetrahedral sites. Compounds like Ga$_2$MgO$_4$, Zn$_2$SnO$_4$, In$_2$MgO$_4$, Mg$_2$SnO$_4$, Al$_2$CdO$_4$ tend to form inverse spinels \cite{Kan_2013, jeyadheepan_preparation_2014, barth_spinel_1932, poix1959etude, hahn_untersuchungen_1955}. Since they are disordered, any experimentally known inverse spinel is omitted from our calculations for this family. Generally, the materials in this family have lower predicted lattice thermal conductivity and Baliga figure of merit than the other two identified families, but there are still quite a few that appear as top performers (see Fig.~\ref{fig:families}). 

Let us first consider the (III)$_2$(II)O$_4$ compounds, also referred to as III-II spinels. Those compounds containing Ga or Al and Mg or Zn are higher performers than their In and Cd counterparts due to larger band gaps. Most of these compounds have been synthesized and are stable at ambient conditions. A few of these materials are of particular interest. Ga$_2$ZnO$_4$ is a known \textit{n}-type single crystal semiconductor as discussed previously \cite{horng_epitaxial_2017, galazka_ultra-wide_2019, chikoidze_p-type_2020}. According to our calculations, it has high Baliga FOM but low $\kappa_L$ and is therefore not a top candidate. Al$_2$MgO$_4$ is a top candidate in our search but is a known insulator used as a single-crystal substrate for semiconductor devices. 

The other top candidate from the III-II spinel family is Al$_2$ZnO$_4$. It has high Baliga FOM, decent $\kappa_L$, and is ranked in the upper half of our \textit{n}-type dopability assessment. Powder samples of stoichiometric, normal Al$_2$ZnO$_4$ have been synthesized, but there are no accounts of single crystal growth, electrical conductivity measurements or attempts to dope this material \cite{levy_structure_2001}. With further investigation this spinel compound may be \textit{n}-type dopable like its gallium counterpart.

The (II)$_2$(IV)O$_4$ compounds have not been considered for power electronics to our knowledge. Like the traditional III-II materials, many of these II-IV versions form in the spinel structure. Depending on the ion species, pressure, and temperature, these compounds can also form in the inverse spinel, phenacite/willemite (s.g. 148), olivine (s.g. 64), modified-spinel/$\beta$ (s.g. 74), and other lower symmetry polymorphs. The tin and cadmium containing compounds have low thermal conductivity and small band gaps (GGA $\mbox{E}_g\,<\,\mbox{2 eV}$) and thus are not of interest for high power electronics.

One top power electronic candidate from the II-IV compounds is Mg$_2$GeO$_4$ spinel. Polycrystals have been synthesized from the olivine ground state structure by heating to 850$^\circ$C in a pressure vessel at 0.5 GPa. Once synthesized, the spinel is stable at ambient conditions, transitioning to olivine above 810$^\circ$C \cite{navrotsky_thermodynamic_1973,von_dreele_refinement_1977}. There is also evidence that this material is intrinsically conductive due to the formation of native donor defects \cite{rehman_growth_2019}. The measured conductivity of above 100 S/cm is larger than that of intrinsic Si, making this an appealing candidate for power electronics. 
The next two candidates with high Baliga FOM and $\kappa_L$ from this group are spinel Mg$_2$SiO$_4$ and trigonal Zn$_2$SiO$_4$. Although the germanate compounds often require less extreme temperature and pressure conditions to form, the silicate analogues have larger band gaps and thus higher Baliga figure of merit as evidenced in Fig.~\ref{fig:families}.

%
Like its germanium counterpart, Mg$_2$SiO$_4$ normal spinel is synthesized from the olivine ground state with an elevated temperature of 1200$^\circ$C and pressure of 22GPa \cite{akaogi_mg2sio4_1984}. Once quenched, the spinel is stable up to about 700$^\circ$C. In our qualitative \textit{n}-type dopabilty analysis, Mg$_2$SiO$_4$ is in the upper third of all candidates but the lowest ranked of the (III/II)$_2$(II/IV)O$_4$ compounds, sitting just below insulator Al$_2$MgO$_4$. Previously published \textit{ab intio} defect calculations indicate that none of the Mg$_2$SiO$_4$ polymorphs are likely \textit{n}-type dopable \cite{verma_ab_2009}. The calculations of olivine structure showed that Mg vacancies and O interstitials will likely compensate any addition electrons from extrinsic dopants. While the study did not conclusively test all possible conditions or defects for the spinel structure, these preliminary results make spinel Mg$_2$SiO$_4$ a less attractive power electronics candidate.

Of all the compounds in the II-IV group, zinc silicate has the most realized stable polymorphs. Select polymorphs are shown in Fig.~\ref{fig:families}. The trigonal (s.g. 122) polymorph is the only Zn$_2$SiO$_4$ with a large enough predicted $\kappa_L$ to be included in the top candidate list. It is the second lowest energy polymorph with a wide range of stability. It can be formed from the ground state "willemite" at 800$^\circ$C and 3-8 GPa \cite{syono_high_1971}. The trigonal version also ranks sixth among the candidates in our \textit{n}-type dopability analysis (see Fig.~\ref{fig:dopability}) which makes it an attractive candidate. To our knowledge, no one has attempted to dope any of the polymorphs of Zn$_2$SiO$_4$. 
Just below our chosen $\kappa_L$ cut-off are two other zinc silicate polymorphs. The willemite structure (s.g. 148) has been grown as a single crystal of 10 mm$^2$ size in previous photo luminescence studies \cite{chang_photoexcited_1981}. The spinel version is just below our top candidate cutoff but, unfortunately, was deemed the rarest and least stable polymorph according to theoretical calculations \cite{karazhanov_phase_2009}. In addition, the ICSD entry for this normal spinel compound is based on theoretical calculations \cite{allali_electronic_2014} and experimental entries all include some degree of inversion. 

Lastly, the HSE06 calculated lattice constants of the four previously unexplored top candidate spinels from this family are all within 2\% of the Al$_2$MgO$_4$ substrate lattice constant as shown in Fig.~\ref{fig:lattice_map}, indicating epitaxial growth on common substrates is possible. If spinel Al$_2$ZnO$_4$ proves \textit{n}-type dopable like some of its other family members, it could be a great power electronic candidate. Mg$_2$GeO$_4$ is also promising considering evidence of intrinsic \textit{n}-type conductivity.  Additionally, given the demonstrated single crystal growth in the willemite Zn$_2$SiO$_4$, its higher energy trigonal polymorph may be a good power electronic candidate as well if a suitable extrinsic dopant is found.  

\subsection{(III)BO$_3$ family}

The last group contains four borates which have (III)BO$_3$ stoichiometry and calcite structure (s.g. 167). Stable polycrystals of all four compounds have been grown. AlBO$_3$ requires hydrothermal high-pressure synthesis techniques \cite{santamar_compressibility_2014} since it does not lie on the convex hull, but once synthesized, remains stable.  In addition, single crystals of InBO$_3$, ScBO$_3$, and GaBO$_3$ have been grown via molten flux at ambient pressures with the GaBO$_3$ crystals measuring 4 $\times$ 4 $\times$ 0.2 mm$^3$ \cite{santamar_compressibility_2014, wang_flux_2015}.  Despite this easy synthesis, to our knowledge no one has attempted to dope these materials or research them for power electronic applications. According to our analysis, these borates rank the lowest for \textit{n}-type dopability potential (see Fig.~\ref{fig:dopability}) among the families. InBO$_3$ sits at about the half-way mark and is the most promising for potential doping of the four. The calcite family has been studied mostly in the pursuit of rare-earth doped phosphors for photoluminescence applications. 

In addition to potential for native substrate growth, there are commercial substrates available for epitaxial growth of heterostructures. ScBO$_3$ and InBO$_3$ lattices match that of the 001 surface of Al$_2$O$_3$ (s.g. 167) within 1.5\% as shown in Fig.~\ref{fig:lattice_map} and Fig.~\ref{fig:lattice_vesta}. Given the very high figures of merit, demonstrated single crystal growth, and potential heteroepitaxial growth, calcites in the (III)BO$_3$ family could make excellent candidates for power electronics.

\subsection{Other Candidates}

There are also a handful of candidate oxides with large Baliga FOM, adequate $\kappa_L$, and small $m^*_{cb}$ that are not part of the families. These deserve to be considered for power electronics as well. We investigate the literature to evaluate the synthesis potential of each of these in the upper half of Fig.~\ref{fig:dopability}. 

Upon further inspection, GeO$_2$ in space group 58 is a high pressure polymorph of rutile (s.g. 136) with very similar properties. The rutile polymorph has already been proposed as a promising WBG semiconductor \cite{chae_rutile_2019, niedermeier_shallow_2020}. Attempts to synthesize thin films of rutile GeO$_2$ have been difficult due to the existence of competing amorphous and quartz phases \cite{chae_epitaxial_2020} and high vapor pressure \cite{chae_toward_2021}, making our high pressure polymorph a less attractive candidate for further exploration. Cd$_2$Ta$_2$O$_7$ is a pyrochlore structure like the identified (III)$_2$(IV)$_2$O$_7$ family but using group II and group V elements. This structure lies on the convex hull and has been grown as a single crystal \cite{kolpakova_low_1992}. For these reasons, Cd$_2$Ta$_2$O$_7$ could be a good power electronic candidate, but the inclusion of a toxic element like cadmium may lower the appeal considering there are safer alternatives. ZnSiO$_3$ (s.g. 148) comes from an ICSD entry based upon a theoretical structure rather than experimental. At 140 meV above the convex hull, this is an unstable polymorph that easily decomposes into SiO$_2$ and the ground state Zn$_2$SiO$_4$ \cite{karazhanov_phase_2009}. Moving down the dopability ranking, CaAl$_4$O$_7$ is a monoclinic structure which lies on the convex hull and has been doped with rare earth elements for luminescence applications \cite{jia_luminescence_2001}. HgTiO$_3$ is a high pressure compound which is metastable at ambient conditions and exhibits ferroelectric behavior \cite{lebedev_ferroelectricity_2012}. Not much more is known about CaAl$_4$O$_7$ or HgTiO$_4$ or their potential for power electronics.

\section{Conclusion}
%

Using a multistage first-principles computational workflow, we conducted a search among binary, ternary, and quaternary oxides for promising wide band gap power electronic semiconductors for high-power, high-voltage, and high-temperature performance. We used a high-throughput evaluation of the Baliga figure of merit and thermal conductivity to down-select candidates based upon theoretical performance followed by a more detailed analysis of the electronic structure, \textit{n}-type dopability, and prior synthesis of the top 40 candidates. Through the use of DFT and phenomenological models, we uphold the expected trends in theoretical performance of currently used or investigated power electronic materials such as Si, GaN, SiC, and $\beta$-Ga$_2$O$_3$. Out of the 1,340 oxides considered, we identified 40 top candidates, most of which are ternaries, which rival the high-power and high-voltage performance of GaN and 4H-SiC, and the lattice thermal conductivity of $\beta$-Ga$_2$O$_3$. 

Of these top materials, many fall into three previously unexplored families of ternary compounds: (III)$_2$(IV)$_2$O$_7$,  (II)$_2$(IV)O$_4$, and (III)BO$_3$. The (III)$_2$(IV)$_2$O$_7$ family contains germanates and silicates that adopt stable thortveitite (s.g. 12) and metstable pyrochlore (s.g. 227) structures. The thortveitite versions have been grown as single crystals which is promising for growth of bulk native substrates. The more symmetric pyrochlore structures will be more difficult to synthesize in bulk, but thin films could be grown heteroepitaxially on common substrates. The Indium-containing members of this family rank high in our qualitative \textit{n}-type dopability analysis. Defect formation energy calculations on pyrochlore In$_2$Ge$_2$O$_7$ confirm that it has the potential to be extrinsically doped without worry of compensation by native acceptor defects if a suitable donor dopant is found. The (II)$_2$(IV)O$_4$ family contains many polymorphs, the most interesting of which is the high symmetry, stable spinel structure. From this set, Mg$_2$GeO$_4$ has exhibited evidence of intrinsic \textit{n}-type conductivity, indicating that it could be easily doped to desired concentrations. In addition, heterostructures of several of these compounds could be grown epitaxially on Al$_2$MgO$_4$ or MgO substrates with acceptable lattice mismatch. The (III)BO$_3$ calcite-structure borates have some of the largest predicted Baliga FOM values and have been grown as single crystals, indicating they could be used as native substrates for vertical power devices. Alternatively, they could be grown heteroepitaxially on Al$_2$O$_3$ substrates. Of these, InBO$_3$ is the most promising candidate as it has the most potential to be \textit{n}-type dopable according to our ranking. Taking into consideration all of these relevant factors, we propose the strongest candidate from each of the identified families for future power electronics research: In$_2$Ge$_2$O$_7$ thortveitite and pyrochlore, Mg$_2$GeO$_4$ spinel, and InBO$_3$ calcite.

Future efforts must address the native defect formation energies of those high-ranked materials that were not confirmed in this study. In addition, suitable extrinsic dopants will need to be determined to realize the proposed materials' potential for power electronics. Once identified, experimental work can confirm these predictions through synthesis, epitaxial growth, and conductivity measurements. These identified wide band gap ternary oxides hold promise for future high-power, high-voltage, and high-temperature power electronics. We hope this work sparks deeper investigation into the potential of these materials to push the performance boundary of power electronics devices and aid in the renewable energy transition.

\begin{acknowledgments}
This work was authored in part at the National Renewable Energy Laboratory (NREL) operated by Alliance for Sustainable Energy, LLC, for the U.S. Department of Energy (DOE) under Contract No. DE-AC36-08GO28308. Funding provided by the Laboratory Directed Research and Development (LDRD) program at NREL and Advanced Energy Systems Graduate Program at the Colorado School of Mines. The research was performed using computational resources sponsored by the DOE’s Office of Energy Efficiency and Renewable Energy located at NREL. The views expressed in the article do not necessarily represent the views of the DOE or the U.S. Government.

\end{acknowledgments}

\section*{SUPPLEMENTAL MATERIAL}
See Supplemental Material for the published data used in the creation of Fig. \ref{fig:motivation} and select calculated properties from this work. 

\nocite{wort_diamond_2008, weingartner_determination_2002, gunning_2012, gunning_2015, zhang_recent_2020, hirama_2020, he_2008, kaplar_2016, nakarmi_2003, allerman_2016, mehnke_2016, baca_2016, sarkar_2018, taniyasu_2006, taniyasu_electrical_2004, ahmad_2021}

\bibliography{PE_oxide_search}

\end{document}


\begin{table}[h]
\caption{\label{tab:fig1_refs} Published measured data for reference power electronic materials and the sources used to create Figure 1. Data includes lattice thermal conductivity ($\kappa$), band gap ($E_g$), bulk critical breakdown field ($E_b$), electron mobility ($\mu_n$) and hole mobility ($\mu_p$) at 10$^{16}$ cm$^{-3}$ carrier concentration, dielectric constant ($\varepsilon$), maximum achieved doping concentration for electrons and holes at room temperature, and largest manufactured wafer diameter. All data without a listed citation is from Kaplar \textit{et al.} [\blue{1}]}

\fontsize{8}{18}\selectfont
\begin{tabular}{C{1.5cm}C{1.1cm}C{1cm}C{1cm}C{1.2cm}C{1.2cm}C{0.8cm}ccc}
\textbf{Compound } & $\bm{\kappa}$ & $\bm{E_g}$ & $\bm{E_b}$ & $\bm{\mu_n}$ & $\bm{\mu_p}$ & $\bm{\varepsilon}$ & \textbf{\textit{n} doping conc.} & \textbf{\textit{p} doping conc.} & \textbf{wafer dia.} [\blue{6}] \\
 & (W/mK) & (eV) & (MV/cm)  & (cm$^2$/Vs) & (cm$^2$/Vs) &  & (cm$^{-3}$) & (cm$^{-3}$) & (mm) \\ \hline
Si   & 150  & 1.1   & 0.3   & 1400  & 480 [\blue{79}]   & 11.9   & 1E+20 [\blue{38}]  & 1E+20 [\blue{38}] & 300 [\blue{3}]  \\
4H-SiC   & 490   & 3.3 & 2.2  & 800  & 120 [\blue{79}] & 10.1 & 1E+19 [\blue{80}]   & 1E+19 [\blue{6}]  & 200  \\
GaN   & 253 [\blue{6}]  & 3.4  & 4.9 [\blue{6}]  & 1000   & 24 [\blue{6}]  & 10.4  & 5E+20 [\blue{38}] & 7.9E+19 [\blue{81}] & 200 \\
$\beta-$Ga$_2$O$_3$  & 27 [\blue{6}]  & 4.9 [\blue{6}]  & 10.3 [\blue{6}] & 184 [\blue{83}]  & unknown   & 10 [\blue{6}]  & 1E+20 [\blue{83}] & unknown  & 100  \\
d-C [\blue{6}]  & 3300   & 5.5  & 13.0  & 4500    & 3800     & 5.7   & 5E+19  & 1E+19 & 25 \\
c-BN  & 768 [\blue{6}]  & 6.4 [\blue{6}] & 17.5 [\blue{6}] & 825 [\blue{6}]  & 500 [\blue{6}]  & 7.1 [\blue{6}]  & 1.2E+18 [\blue{84}]  & 6E+18 [\blue{85}]  & 5  \\
Al$_{0.3}$Ga$_{0.7}$N  & 40 & 4.1  & 5.9  & 150 [\blue{86}] & 3 [\blue{87}]  & 10.3  & 7E+18 [\blue{43}] & 2.3E+19 [\blue{88}] & 200 \\
Al$_{0.85}$Ga$_{0.15}$N & 50 & 5.7  & 13.4   & 250 [\blue{89}] & unknown  & 10.2  & 2.7E+18 [\blue{45}] & unknown & 50 \\
AlN  & 340 [\blue{42}] & 6.2 & 15.4 [\blue{6}]  & 425 [\blue{90}]  & 2 [\blue{91}] & 10.1  & 1.75E+15 [\blue{91}] & 3.1E+18 [\blue{92}] & 50                        
\end{tabular}

\end{table}

\fontsize{8}{18}\selectfont
\begin{longtable}[c]{C{1.2cm}C{0.4cm}|C{0.5cm}C{0.6cm}C{0.9cm}C{0.8cm}C{1.05cm}C{1.1cm}|C{0.5cm}C{0.6cm}C{0.9cm}C{0.8cm}C{1.05cm}C{1.1cm}}
\caption{\label{tab:GGAvsHSE} A comparison of computed properties from electronic structure calculations of the top 40 \textit{n}-type power electronic candidates using GGA+U or HSE06 methods. This includes the international space group number (s.g.), the band gap ($E_g$), band effective mass for electrons ($m^*_{b,cb}$), electron mobility ($\mu_n$), breakdown field ($E_b$), and \textit{p}-type and \textit{n}-type Baliga figure of merit. The Baliga FOM has been normalized to the calculated \textit{n}-type Silicon Baliga FOM using GGA or HSE06 methods. Candidates are listed in alphabetical order by element. Related to Table 1, Figure 3, and Figure 5.}

 \\\multicolumn{2}{c}{ }& \multicolumn{6}{c}{GGA+U}& \multicolumn{6}{c}{HSE06}\\
 \textbf{Compound} & \textbf{s.g.}  & \bm{$E_g$} & \bm{$m^*_{b,cb}$} &\bm{ $\mu_n$ }& \bm{$E_b$} & \textbf{BFOM(p)}  & \textbf{BFOM(n)}  & \bm{$E_g$ }& \bm{$m^*_{b,cb}$} &\bm{ $\mu_n$ }& \bm{$E_b$}    & \textbf{BFOM(p)}  & \textbf{BFOM(n)} \\ 
  &   & (eV) & ($m_{e}$) & (cm$^2$/Vs) & (MV/cm) &  p-norm & n-norm & (eV) & ($m_{e}$) & (cm$^2$/Vs) & (MV/cm)    & p-norm  &n-norm \\ \hline
AgTaO$_3$                    & 161 & 2.68 & 1.19 & 14.5   & 3.6  & 90   & 588   & 3.40 & 0.31 & 106.5  & 5.0  & 39   & 3424    \\
AgTaO$_3$                    & 167 & 2.68 & 1.17 & 16.1   & 3.6  & 99   & 653   & 3.40 & 0.31 & 118.3  & 5.0  & 45   & 3806    \\
Al$_2$MgO$_4$             & 227 & 5.17 & 0.10 & 691.9  & 7.1  & 13   & 4419  & 7.24 & 0.10 & 681.3  & 13.0 & 30   & 7929    \\
Al$_2$O$_3$               & 167 & 5.88 & 0.10 & 698.6  & 8.5  & 53   & 9833  & 8.30 & 0.10 & 860.9  & 16.6 & 240  & 26087   \\
Al$_2$ZnO$_4$             & 227 & 3.88 & 0.09 & 856.2  & 4.7  & 8    & 1661  & 5.98 & 0.10 & 758.3  & 9.5  & 22   & 3586    \\
Al$_3$B$_4$YO$_{12}$ & 155 & 5.64 & 2.50 & 5.4    & 27.1 & 128  & 2343  & 7.34 & 3.79 & 2.5    & 52.6 & 670  & 2313    \\
AlTaO$_4$                    & 60  & 3.56 & 0.73 & 45.7   & 4.9  & 235  & 370   & 4.96 & 0.76 & 43.1   & 8.4  & 491  & 512     \\
BAlO$_3$                     & 167 & 5.74 & 0.60 & 51.2   & 24.4 & 2222 & 13257 & 7.83 & 0.65 & 45.3   & 52.8 & 7960 & 34432   \\
C$_2$La$_2$O$_2$       & 12  & 2.90 & 2.27 & 2.7    & 13.8 & 198  & 224   & 4.43 & 2.81 & 1.9    & 35.5 & 693  & 799     \\
Ca$_2$Ta$_2$O$_7$      & 227 & 2.52 & 0.51 & 48.7   & 3.2  & 20   & 201   & 3.71 & 0.41 & 66.7   & 5.7  & 33   & 419     \\
CaAl$_4$O$_7$             & 15  & 4.03 & 0.14 & 269.0  & 6.6  & 10   & 1558  & 5.84 & 0.14 & 258.8  & 13.0 & 24   & 3292    \\
Cd$_2$Ta$_2$O$_7$      & 227 & 2.04 & 0.07 & 919.3  & 2.5  & 4    & 3525  & 3.66 & 0.09 & 715.6  & 5.6  & 15   & 8559    \\
CuTaO$_3$                    & 161 & 1.75 & 0.63 & 50.4   & 1.9  & 14   & 353   & 2.09 & 0.46 & 82.7   & 2.3  & 11   & 298     \\
GaBO$_3$                     & 167 & 3.84 & 0.25 & 173.4  & 10.3 & 145  & 4229  & 6.30 & 0.29 & 134.9  & 29.6 & 2317 & 22240   \\
GeHfO$_4$                    & 88  & 3.99 & 0.12 & 622.9  & 4.7  & 6    & 2375  & 6.31 & 0.81 & 34.3   & 10.2 & 32   & 372     \\
GeO$_2$                      & 58  & 1.25 & 0.04 & 3140.3 & 1.3  & 1.4  & 228   & 3.87 & 0.05 & 1940.2 & 5.0  & 27   & 2206    \\
GeO$_2$                      & 136 & 1.21 & 0.04 & 3158.8 & 1.3  & 1.5  & 230   & 3.85 & 0.05 & 1962.3 & 4.9  & 27   & 2168    \\
HfB$_2$O$_5$              & 14  & 4.74 & 1.43 & 15.2   & 13.7 & 86   & 909   & 6.29 & 1.84 & 10.4   & 25.2 & 509  & 1124    \\
HfSiO$_4$                    & 141 & 5.69 & 4.66 & 3.7    & 14.5 & 373  & 273   & 7.25 & 2.82 & 5.2    & 24.4 & 617  & 532     \\
In$_2$Ge$_2$O$_7$      & 227 & 1.11 & 0.03 & 5385.0 & 1.2  & 0.2  & 410   & 3.21 & 0.04 & 3043.5 & 3.7  & 1    & 1907    \\
In$_2$Ge$_2$O$_7$      & 12  & 1.82 & 0.04 & 1765.7 & 2.3  & 0.5  & 706   & 3.96 & 0.05 & 1252.8 & 6.7  & 3    & 3522    \\
In$_2$Si$_2$O$_7$      & 227 & 1.84 & 0.05 & 2398.5 & 2.0  & 0.5  & 787   & 3.82 & 0.06 & 1909.8 & 5.1  & 6    & 2932    \\
In$_2$Si$_2$O$_7$      & 12  & 2.74 & 0.09 & 659.9  & 5.0  & 3    & 1637  & 4.81 & 0.10 & 557.7  & 13.3 & 34   & 7485    \\
InBO$_3$                     & 167 & 2.71 & 0.15 & 283.3  & 5.6  & 3    & 1028  & 4.80 & 0.16 & 278.6  & 15.7 & 104  & 6477    \\
Mg$_2$GeO$_4$             & 227 & 3.08 & 0.08 & 796.8  & 3.2  & 2    & 479   & 5.32 & 0.09 & 696.8  & 7.1  & 7    & 1347    \\
Mg$_2$SiO$_4$             & 227 & 4.82 & 0.10 & 632.0  & 7.4  & 5    & 4247  & 7.03 & 0.11 & 599.0  & 15.0 & 38   & 9667    \\
MgScBO$_4$                   & 62  & 4.60 & 0.98 & 16.1   & 17.8 & 208  & 2226  & 6.38 & 1.14 & 12.8   & 38.0 & 1145 & 4967    \\
Sc$_2$Si$_2$O$_7$      & 12  & 4.70 & 1.89 & 7.2    & 12.8 & 55   & 345   & 6.39 & 1.57 & 9.5    & 24.9 & 144  & 969     \\
ScBO$_3$                     & 167 & 4.71 & 1.41 & 10.9   & 15.0 & 72   & 1094  & 6.34 & 1.15 & 15.0   & 29.1 & 170  & 3146    \\
SrZrO$_3$                    & 62  & 4.06 & 0.41 & 97.1   & 4.9  & 11   & 693   & 5.24 & 0.30 & 100.7  & 7.4  & 13   & 702  \\
Ta$_2$O$_5$               & 1   & 2.48 & 0.49 & 63.3   & 3.0  & 35   & 214   & 3.69 & 0.30 & 134.5  & 5.3  & 64   & 684     \\
TiHgO$_3$                    & 161 & 1.37 & 0.16 & 334.9  & 1.4  & 4    & 281   & 2.82 & 0.15 & 350.9  & 2.8  & 13   & 781     \\
Y$_2$Si$_2$O$_7$       & 12  & 4.66 & 0.18 & 223.5  & 11.0 & 37   & 5356  & 6.58 & 0.21 & 171.3  & 22.7 & 112  & 10311   \\
Y$_2$Si$_2$O$_7$       & 14  & 4.76 & 0.16 & 232.6  & 12.8 & 262  & 8975  & 6.71 & 0.19 & 186.5  & 26.7 & 1522 & 19042   \\
Zn$_2$SiO$_4$             & 122 & 2.66 & 0.07 & 802.9  & 3.5  & 0.6  & 705   & 4.68 & 0.08 & 689.1  & 8.4  & 3    & 2351    \\
Zn$_4$B$_6$O$_{13}$  & 217 & 3.43 & 0.19 & 269.1  & 6.7  & 7    & 1274  & 5.42 & 0.19 & 257.7  & 15.7 & 30   & 4516    \\
ZnB$_4$O$_7$              & 63  & 5.42 & 0.22 & 250.1  & 18.8 & 1913 & 27681 & 7.74 & 0.20 & 297.1  & 44.0 & 7433 & 121329  \\
ZnCO$_3$                     & 167 & 3.50 & 0.49 & 38.5   & 11.8 & 187  & 1297  & 5.66 & 0.44 & 44.7   & 33.9 & 1345 & 10217   \\
ZnSiO$_3$                    & 148 & 3.61 & 0.09 & 870.5  & 4.2  & 5    & 1863  & 5.92 & 0.10 & 685.4  & 9.3  & 30   & 4620    \\
ZrB$_2$O$_5$              & 14  & 4.15 & 2.08 & 8.4    & 10.3 & 49   & 226   & 5.63 & 2.24 & 7.5    & 19.1 & 293  & 371    
\end{longtable}

\newpage

\newpage
\begin{table}[ht!]
\caption{\label{tab:GGAvsHSE-refs} A comparison of computed properties from electronic structure calculations of the reference materials using GGA+U or HSE06 methods. First set of reference materials is current power electronic semiconductors and is related to Figure 3 and Figure 4. Second set of reference materials is known \textit{n}-type dopable or \textit{n}-type insulating materials and is related to Figure 5. This includes the international space group number (s.g.), the band gap ($E_g$), band effective mass for electrons ($m^*_{b,cb}$), electron mobility ($\mu_n$), breakdown field ($E_b$), and \textit{p}-type and \textit{n}-type Baliga figure of merit. The Baliga FOM has been normalized to the calculated \textit{n}-type Silicon Baliga FOM using GGA or HSE06 methods. Materials are listed in alphabetical order.}

\fontsize{8}{18}\selectfont
\begin{tabular}[c]{C{1.2cm}C{0.4cm}|C{0.5cm}C{0.6cm}C{0.9cm}C{0.8cm}C{1.05cm}C{1.1cm}|C{0.5cm}C{0.6cm}C{0.9cm}C{0.8cm}C{1.05cm}C{1.1cm}}

 \multicolumn{2}{c}{}& \multicolumn{6}{c}{GGA+U}& \multicolumn{6}{c}{HSE06} \\
 \textbf{Compound} & \textbf{s.g.}  & \bm{$E_g$} & \bm{$m^*_{b,cb}$} &\bm{ $\mu_n$ }& \bm{$E_b$} & \textbf{BFOM(p)}  & \textbf{BFOM(n)}  & \bm{$E_g$ }& \bm{$m^*_{b,cb}$} &\bm{ $\mu_n$ }& \bm{$E_b$}    & \textbf{BFOM(p)}  & \textbf{BFOM(n)} \\ 
  &   & (eV) & ($m_{e}$) & (cm$^2$/Vs) & (MV/cm) &  p-norm & n-norm & (eV) & ($m_{e}$) & (cm$^2$/Vs) & (MV/cm)    & p-norm  &n-norm \\ \hline

\multicolumn{14}{c}{\textbf{Power Electronic References}}   \\                          \hline                                                                            
C                 & 227 & 4.14 & 0.15 & 898.1  & 13.1 & 27200 & 25846 & 5.36 & 0.12 & 1241.8 & 22.8 & 53859 & 54186  \\
Ga$_2$O$_3$ & 12  & 2.01 & 0.04 & 2292.9 & 1.9  & 0.2   & 419   & 4.22 & 0.05 & 1528.7 & 4.8  & 1.3   & 1276   \\
GaN               & 186 & 1.76 & 0.03 & 3913.8 & 1.6  & 2     & 355   & 3.29 & 0.04 & 2710.8 & 3.1  & 5     & 545    \\
Si                & 227 & 0.62 & 0.19 & 124.7  & 0.6  & 0.5   & 1     & 1.16 & 0.17 & 150.2  & 0.9  & 0.5   & 1      \\
4H-SiC           & 186 & 2.23 & 0.12 & 606.6  & 2.8  & 18    & 318   & 3.18 & 0.11 & 719.3  & 4.6  & 24    & 449    \\
6H-SiC         & 186 & 2.04 & 0.25 & 76.3   & 2.6  & 8     & 31    & 2.95 & 0.19 & 315.3  & 4.2  & 16    & 159    \\
8H-SiC            & 186 & 1.78 & 0.16 & 428.7  & 2.3  & 7     & 111   & 2.69 & 0.14 & 503.5  & 3.7  & 10    & 172    \\
\hline
\multicolumn{14}{c}{\textbf{\textit{N}-type Dopability   References}}        \\                 \hline                                                                            
CaZrSi$_2$O$_7$ & 5   & 4.88 & 3.36 & 2.6    & 11.4 & 36     & 89  & 6.45 & 3.97 & 2.0    & 20.3 & 71     & 112   \\
Cu$_2$O            & 224 & 0.71 & 0.33 & 64.6   & 0.8  & 0.1    & 1.0 & 1.95 & 0.40 & 47.0   & 1.5  & 0.2    & 1.5   \\
CuInO$_2$          & 166 & 0.48 & 0.14 & 267.7  & 0.6  & 0.02   & 1.3 & 1.73 & 0.18 & 171.6  & 1.5  & 0.1    & 3     \\
Ga$_2$MgO$_4$   & 227 & 3.15 & 0.07 & 1212.5 & 3.0  & 0.6    & 709 & 5.18 & 0.07 & 1003.4 & 5.9  & 0.6    & 1379  \\
Ga$_2$ZnO$_4$   & 227 & 2.33 & 0.05 & 2061.3 & 2.1  & 0.3    & 465 & 4.39 & 0.06 & 1374.5 & 4.6  & 0.9    & 970   \\
In$_2$O$_3$     & 206 & 0.97 & 0.03 & 3848.1 & 0.9  & 0.03   & 68  & 2.67 & 0.04 & 2233.2 & 0.6  & 0.001  & 3     \\
ScAlO$_3$          & 62  & 4.95 & 2.32 & 6.9    & 6.0  & 94     & 63  & 6.59 & 2.95 & 4.8    & 9.7  & 141    & 54    \\
SiO$_2$            & 154 & 5.72 & 0.19 & 39.9   & 0.5  & 0.0002 & 0.1 & 7.73 & 0.18 & 42.5   & 0.6  & 0.0001 & 0.03  \\
SnO$_2$            & 136 & 0.69 & 0.03 & 3871.4 & 0.8  & 0.2    & 58  & 2.94 & 0.04 & 2214.9 & 3.0  & 3      & 452   \\
TiO$_2$            & 136 & 2.07 & 0.99 & 24.4   & 2.2  & 14     & 34  & 3.33 & 0.60 & 51.6   & 4.0  & 23     & 122   \\
TiO$_2$            & 141 & 2.38 & 0.56 & 50.3   & 2.0  & 22     & 27  & 3.62 & 0.41 & 76.0   & 3.3  & 35     & 52    \\
ZnO                   & 225 & 0.73 & 0.03 & 3859.6 & 0.5  & 0.01   & 26  & 2.66 & 0.04 & 2341.5 & 1.0  & 0.03   & 32    \\
ZnO                   & 186 & 0.74 & 0.03 & 2863.3 & 0.7  & 0.1    & 22  & 2.48 & 0.05 & 1583.1 & 1.7  & 0.4    & 52    \\
ZrO$_2$            & 14  & 3.79 & 2.34 & 5.7    & 0.9  & 0.1    & 0.1 & 5.11 & 2.08 & 6.8    & 1.1  & 0.04   & 0.1   \\
\end{tabular}

\end{table}

\newpage

\begin{table}[ht!]
\caption{\label{tab:HSE_bulk} Comparison of bulk modulus ($B$) calculations for select compounds using GGA+U and HSE06 methods. Included as well are the calculated normalized \textit{n}-type Baliga figures of merit (BFOM) using the GGA+U electronic structure + GGA+U bulk modulus, HSE06 electronic structure + GGA+U bulk modulus, and HSE06 electronic structure + HSE06 bulk modulus. Results show that calculating the electronic structure using HSE06 increases the BFOM by an order of magnitude in most cases. Also, including the HSE06 calculated bulk modulus only increases the BFOM by a modest amount in comparison.}
\fontsize{8}{18}\selectfont
\begin{tabular}{ccccC{2.8cm}C{2.5cm}C{2.5cm}}
  & & & & \textbf{GGA+U} & \textbf{HSE06} & \textbf{HSE06}\\
\textbf{Compound}           & \textbf{s.g.}  & \textbf{B (GGA+U) }& \textbf{B (HSE06)} & \textbf{BFOM$_{\mbox{n-norm}}$}  & \textbf{BFOM$_{\mbox{n-norm}}$}  & \textbf{BFOM$_{\mbox{n-norm}}$}  \\
 & & (GPa) & (GPa) &  \textbf{with GGA+U B}& \textbf{with GGA+U B} & \textbf{with HSE06 B}\\

\hline
Al$_2$MgO$_4$      & 227 & 183        & 204       & 4.42E+03   & 7.93E+03      & 8.85E+03 \\
GaN                & 186 & 171        & 198       & 3.55E+02   & 5.45E+02      & 6.31E+02  \\
GeHfO$_4$          & 88  & 209        & 238       & 2.37E+03    & 3.72E+02  & 4.23E+02    \\
HfB$_2$O$_5$       & 14  & 217        & 238       & 9.09E+02  & 1.12E+03    & 1.23E+03   \\
In$_2$Si$_2$O$_7$  & 12  & 147        & 169       & 1.64E+03   & 7.49E+03    & 8.56E+03  \\
Sc$_2$Si$_2$O$_7$  & 12  & 156        & 170       & 3.45E+02 & 9.69E+02    & 1.06E+03   \\
ScBO$_3$           & 167 & 154        & 167       & 1.09E+03 & 3.15E+03   & 3.43E+03   \\
Y$_2$Si$_2$O$_7$   & 12  & 134        & 145       & 5.36E+03   & 1.03E+04  & 1.12E+04   \\
Al$_3$B$_4$YO$_{12}$ & 155 & 178        & 196       & 2.34E+03   & 2.65E+03   & 2.93E+03  \\
Zn$_2$SiO$_4$      & 122 & 132        & 144       & 7.05E+02 & 2.35E+03  & 2.57E+03   \\
Zn$_4$B$_6$O$_{13}$  & 217 & 182        & 200       & 1.27E+03   & 4.52E+03   & 4.95E+03  \\
ZnB$_4$O$_7$       & 63  & 214        & 238       & 2.77E+04    & 1.21E+05  & 1.34E+05   \\
ZnSiO$_3$          & 148 & 173        & 204       & 1.86E+03  & 4.62E+03  & 5.44E+03 \\

\end{tabular}
\end{table}